\documentclass[]{aa}
\usepackage{graphicx}
\graphicspath{{./figures/}}
\usepackage{natbib}
\usepackage{amssymb,amsmath,amsfonts}
\usepackage{epsfig}
\usepackage{mathptmx}
\usepackage{longtable}
\usepackage{comment}
\usepackage{color}
\usepackage[caption=false]{subfig}
\usepackage{animate}
\usepackage{rotating}
\usepackage{pdflscape}
\usepackage{url}
\usepackage{enumitem}
\usepackage{hyperref}
\usepackage[normalem]{ulem}
\usepackage{longtable} 
\usepackage{pdflscape} 

\begin{document} 
\title{The LOFAR Two-metre Sky Survey: Deep Fields Data Release 2}
\subtitle{I. The ELAIS-N1 field} 
\authorrunning{Shimwell et~al.}
\titlerunning{ELAIS-N1}
\author{T. W. Shimwell$^{\ref{ASTRON},\ref{Leiden}}$\thanks{E-mail: shimwell@astron.nl},
C. L. Hale$^{\ref{Oxford},\ref{Edinburgh}}$,
P. N. Best$^{\ref{Edinburgh}}$, 
A. Botteon$^{\ref{INAF}}$, 
A. Drabent$^{\ref{Tautenburg}}$, 
M. J. Hardcastle$^{\ref{Hertfordshire}}$, 
V. Jeli\'c$^{\ref{Zagreb}}$,
J. M. G. H. J. de Jong$^{\ref{Leiden}}$,
R. Kondapally$^{\ref{Edinburgh}}$,
H. J. A. R\"{o}ttgering$^{\ref{Leiden}}$,
C. Tasse$^{\ref{GEPI},\ref{Rhodes}}$,
R. J. van Weeren$^{\ref{Leiden}}$,
W. L. Williams$^{\ref{SKA}}$,
A. Bonafede$^{\ref{Bologna},\ref{INAF}}$,
M. Bondi$^{\ref{INAF}}$,
M. Br\"uggen$^{\ref{Hamburg}}$,
G. Brunetti$^{\ref{INAF}}$, 
J. R. Callingham$^{\ref{ASTRON},\ref{Leiden}}$,
F. De Gasperin$^{\ref{INAF}}$, 
K. J. Duncan$^{\ref{Edinburgh}}$, 
C. Horellou$^{\ref{Chalmers}}$,
S. Iyer$^{\ref{Amsterdam}}$,
I. de Ruiter$^{\ref{Amsterdam}}$,
K. Ma\l{}ek$^{\ref{Warsaw}}$,
D. G. Nair$^{\ref{Concepción},\ref{MaxPlanck}}$,
L. K. Morabito$^{\ref{Durham1},\ref{Durham2}}$,
I. Prandoni$^{\ref{INAF}}$,
A. Rowlinson$^{\ref{Amsterdam},\ref{ASTRON}}$,
J. Sabater,
A. Shulevski$^{\ref{ASTRON},\ref{Groningen},\ref{Amsterdam}}$,
D. J. B. Smith$^{\ref{Hertfordshire}}$,
F. Sweijen$^{\ref{Durham1}}$
}

\institute{ASTRON, the Netherlands Institute for Radio Astronomy, Postbus 2, NL-7990 AA Dwingeloo, The Netherlands \label{ASTRON} \and
Leiden Observatory, Leiden University, PO Box 9513, 2300 RA Leiden, The Netherlands \label{Leiden} \and
Astrophysics, University of Oxford, Denys Wilkinson Building, Keble Road, Oxford, OX1 3RH, UK \label{Oxford} \and
Institute for Astronomy, School of Physics and Astronomy, University of Edinburgh, Royal Observatory, Blackford Hill, Edinburgh, EH9 3HJ, UK \label{Edinburgh}
\and
INAF-Istituto di Radioastronomia, Via P. Gobetti 101, 40129 Bologna, Italy \label{INAF}
\and
Th\"uringer Landessternwarte, Sternwarte 5, D-07778 Tautenburg, Germany \label{Tautenburg}
\and
Centre for Astrophysics Research, University of Hertfordshire, Hatfield, Herts, AL10 9AB, UK	\label{Hertfordshire} 
\and
Ru{\dj}er Bo\v{s}kovi\'c Institute, Bijeni\v{c}ka cesta 54, 10000 Zagreb, Croatia \label{Zagreb}
\and
GEPI, Observatoire de Paris, Universit\'{e} PSL, CNRS,  5 Place
Jules Janssen, 92190 Meudon, France \label{GEPI}
\and
Department of Physics \& Electronics, Rhodes University, PO Box 94, Grahamstown, 6140, South Africa \label{Rhodes}
\and
SKA Observatory, Jodrell Bank, Lower Withington, Macclesfield, SK11 9FT, United Kingdom	\label{SKA}
\and
DIFA - Universit\`a di Bologna, via P. Gobetti 93, 40129, Bologna \label{Bologna} 
\and
Hamburg Observatory, University of Hamburg, Gojenbergsweg 112, 21029 Hamburg, Germany	\label{Hamburg}
\and
 Chalmers University of Technology, Onsala Space Observatory, SE-43992 Onsala, Sweden
 \label{Chalmers}
\and
Anton Pannekoek Institute for Astronomy, University of Amsterdam, Postbus 94249, 1090 GE Amsterdam, The Netherlands \label{Amsterdam}
\and
National Centre for Nuclear Research, Pasteura 7, 02-093 Warsaw, Poland \label{Warsaw}
\and
Astronomy Department, Universidad de Concepción, Casilla 160-C, Concepción, Chile 
\label{Concepción}
\and
Max Planck Institute for Radio Astronomy,
Auf dem H{\"u}gel 69, 53121 Bonn, Germany
\label{MaxPlanck}
\and
Centre for Extragalactic Astronomy, Department of Physics, Durham University, South Road, Durham DH1 3LE, UK \label{Durham1} 
\and
Institute for Computational Cosmology, Department of Physics, Durham University, South Road, Durham DH1 3LE, UK \label{Durham2} 
\and
Kapteyn Astronomical Institute, University of Groningen, Landleven 12, 9747 AD Groningen, The Netherlands \label{Groningen}
}

\date{Accepted ---; received ---; in original form \today}

\abstract{
\noindent
We present the final 6$\arcsec$ resolution data release of the ELAIS-N1 field from the LOw-Frequency ARray (LOFAR) Two-metre Sky Survey Deep Fields project (LoTSS Deep).  The 144\,MHz images are the most sensitive achieved to date at this frequency and were created from 290\,TB of data obtained from 505\,hrs on-source observations taken over 7.5 years. The data were processed following the strategies developed for previous LoTSS and LoTSS Deep data releases. The resulting images span 24.53 square degrees and, using a refined source detection approach, we identified 154,952 radio sources formed from 182,184 Gaussian components within this area. The maps reach a noise level of 10.7\,$\mu$Jy beam$^{-1}$ at 6$\arcsec$ resolution where approximately half of the noise is due to source confusion. In about 7.4\% of the image our limited dynamic range around bright sources results in a further $>5\%$ increase in the noise. The images have a flux density scale accuracy of about 9\% and the standard deviation of offsets between our source positions and those from Pan-STARRS is 0.2$\arcsec$ in RA and Dec for high significance detections. We searched individual epoch images for variable sources, identifying 39 objects with considerable variation. We also searched for circularly polarised sources achieving three detections of previously known emitters (two stars and one pulsar) whilst constraining the typical polarisation fraction plus leakage to be less than 0.045\%.}
 \maketitle

\section{Introduction}
 
The European Large Area ISO Survey-North 1 (ELAIS-N1; \citealt{Oliver_2000}) field is one of the most widely studied extragalactic fields that can be observed by the LOw-Frequency ARray (LOFAR; \citealt{vanHaarlem_2013}) at high elevations, corresponding to optimal sensitivity. It has a wealth of data at radio wavelengths (see e.g. Sect. 2.1 of \citealt{Sabater_2021}) and across the rest of the electromagnetic spectrum (see e.g. Sect. 2.3.1 of \citealt{Best_2023}). This extensive dataset allows for the classification of radio sources and the determination of their physical properties. Multi-wavelength efforts like this have revealed that the vast majority of detectable extragalactic radio sources are star-forming galaxies and active galactic nuclei, AGN (e.g. \citealt{Smolcic_2017}, \citealt{Whittam_2022},  \citealt{Best_2023} and \citealt{Das_2024}). Using deep multi-epoch observations from the most sensitive operating radio interferometers, the evolution of these source populations can be probed out to high redshift ($z>4$) and aspects such as the time variability or polarisation properties of any detected emission can be examined (e.g. \citealt{Kondapally_2021}, \citealt{Duncan_2021}, \citealt{Callingham_2021b}, \citealt{Smolcic_2017b}, \citealt{Whittam_2022}, \citealt{Kondapally_2022}, \citealt{Cochrane_2023} and \citealt{Piras_2024}).

ELAIS-N1 together with three other well-studied extragalactic fields --- Bo{\"o}tes (\citealt{Jannuzi_1999}), the Lockman Hole (\citealt{Lockman_1986}), and the Euclid Deep Field North (\citealt{Euclid_2022}) which covers the North Ecliptic Pole (NEP) --- form the LOFAR Two-metre Sky Survey Deep Fields (LoTSS Deep; see \citealt{Best_2023}). Each of these fields has been targeted with between 376 and 505\,hrs of LOFAR High Band Antenna (HBA) observations aiming to reach RMS noise levels of order 10-20\,$\mu$Jy beam$^{-1}$ at 120-168MHz when imaging at an angular resolution of 6$\arcsec$. Initial images of the fields reaching RMS noise levels as low as 17.1\,$\mu$Jy beam$^{-1}$ and using up to 163.7\,hrs of data have already been published in the first data release from LoTSS Deep (\citealt{Sabater_2021}, \citealt{Tasse_2021}, \citealt{Bondi_2024}). These studies were accompanied by descriptions of the techniques used to produce the radio maps (\citealt{Tasse_2021}) and for all fields, except the NEP which was processed later, multi-wavelength catalogues (\citealt{Kondapally_2021}), redshifts (\citealt{Duncan_2021}) and host galaxy properties have been derived (\citealt{Best_2023} and \citealt{Das_2024}). LoTSS Deep has already facilitated many studies in  areas such as 
radio source populations (e.g. \citealt{Mandal_2021}), 
AGN (e.g. \citealt{Mingo_2022}, \citealt{Kondapally_2022}, \citealt{Kondapally_2023}, \citealt{Calistro_Rivera_2023} and \citealt{deJong_2024})
star formation (e.g. \citealt{Bonato_2021}, \citealt{Smith_2021}, \citealt{Cochrane_2023}),
cosmology (e.g. \citealt{Hardcastle_2021}, \citealt{Gloudemans_2021}),
galaxy clusters (e.g. \citealt{Osinga_2021}),
magnetic fields (e.g. \citealt{Piras_2024}) and even flare stars (e.g. \citealt{Callingham_2021b}).

Excitingly, the data are also being used to probe the low frequency sky at sub-arcsecond resolution by making use of the international LOFAR stations (e.g. \citealt{Morabito_2022}), which were not included in the first LoTSS Deep data release due to data processing limitations but for which data are present in the majority of the observations taken. Including these stations provides significant additional collecting area and sensitivity to compact sources whilst allowing for confusion noise limitations to be mitigated. The first high resolution wide-field studies using the LoTSS Deep data have recently been released with \cite{Sweijen_2022} mapping 8\,hrs of Lockman Hole data, \cite{deJong_2024} mapping 32\,hrs of ELAIS-N1 data (RMS of 14\,$\mu$Jy beam$^{-1}$) and upcoming studies by Escott et al. (in prep) and de Jong et al. (in prep) using 8\,hrs and at least 216\,hrs of Bootes and ELAIS-N1 data respectively. These very high resolution images have lower surface brightness sensitivity but are complemented by intermediate ($\sim1.2\arcsec$) resolution images such as those from \cite{Ye_2023} and \cite{deJong_2024} to allow the detection of emission from more diffuse objects. Another limitation of LoTSS Deep is the lack of accurate radio spectral information for many of the detected sources, particularly at lower frequencies. Towards rectifying this, a first study by \cite{Williams_2021} has been carried out with the LOFAR Low Band Antenna (LBA) and there are plans for deeper LBA studies in the future potentially reaching frequencies as low as 10\,MHz (e.g. \citealt{Groeneveld_2024}). Finally, the fields are also being mapped in linear polarisation with 
\cite{Ruiz_2021} and \cite{Piras_2024} searching deep ELAIS-N1 datasets (48 and 176\,hrs respectively) for emission from discrete radio sources, whilst \cite{Jelic_2014} and \cite{Snidaric_2023} have examined diffuse polarised emission from our own Galaxy using 8\,hrs and 150\,hrs of ELAIS-N1 data respectively.

Here we release the final 6$\arcsec$ resolution LoTSS Deep ELAIS-N1 image. In this study we use all 505\,hrs of data taken for the LoTSS Deep ELAIS-N1 project to form a very deep synthesized image with an RMS noise level of 10.7\,$\mu$Jy beam$^{-1}$ at 144\,MHz. This is the most sensitive map to date at this frequency. A significant ($\approx 50\%$) fraction of the image noise is from source confusion, implying that this image is approaching the sensitivity limit at 6$\arcsec$ resolution. We use the deep 6$\arcsec$ resolution map to characterise image properties including the flux density scale, astrometric accuracy, dynamic range and the variation between individual epoch images. We also perform analyses to examine the level of confusion noise, quantify the variability of sources in the field between epochs, and search for circularly polarised emission (linear polarisation will be a future study) and for very diffuse emission in the field.

LoTSS Deep is one of several efforts to approximately match the $\mu$Jy beam$^{-1}$ sensitivity levels at GHz frequencies of the deepest  radio surveys (e.g. \citealt{Smolcic_2017}, \citealt{Vlugt_2021} and \citealt{Heywood_2021}) but over a larger area of sky (tens of square degrees) to obtain more representative samples of sources and include rarer objects. The MeerKAT International GigaHertz Tiered Extragalactic Exploration (MIGHTEE; \citealt{Jarvis_2016}) survey has similar aims but at GHz frequencies and the first images from that survey have already been released (\citealt{Heywood_2022} and \citealt{Hale_2024}). Both LoTSS Deep at 6$\arcsec$ and MIGHTEE at $5\arcsec$ reach comparable depths, as both are severely limited in the deepest regions of the image by confusion noise. This limitation further highlights the need for higher resolution wide-field studies such as those by \cite{Sweijen_2022}, \cite{deJong_2024}, \cite{Ye_2023} and \cite{Morabito_2022} which can potentially probe far deeper by mitigating confusion noise.

In Sect. \ref{obs_and_cal} we provide an overview of the observations, data processing and cataloguing. In Sect. \ref{Sec:image_quality} we examine the quality of the ELAIS-N1 deep image. In Sect. \ref{sec:image_analysis} we analyse selected aspects of the dataset to demonstrate its scientific potential and limitations. In Sect. \ref{sec:data_release} we describe the data release products before summarising in Sect. \ref{sec:summary}

\section{Observations and data calibration}
\label{obs_and_cal}

The ELAIS-N1 field has been observed with the LOFAR HBA `dual inner' observing mode for a total of 505\,hrs, consisting of 64 individual exposures each using between 54 and 75 independent stations taken over a period of 7.5 years (from 2014 May 19 to 2021 Oct 30) through four different LoTSS Deep observing campaigns. In this observing mode the core HBA sub-stations each act as an independent station and only the inner 24 tiles of the remote stations are used to match the area of the core HBA sub-stations to make the primary beam the same for all stations in the Netherlands. Each observation is bookended by 10\,min scans of a calibrator source (always 3C295 and generally also 3C147). An overview of the dataset and resulting images is given in Tab. \ref{LOFAR-obs} and details of the individual observations are provided in \href{https://doi.org/10.5281/zenodo.14603969}{Tab. S1}. We note that here we have already excluded 24\,hrs of LOFAR cycle 0 observations (project LC0\_019) which had poor noise levels or data issues (see \citealt{Sabater_2021} for further details) and 47.4\,hrs of LOFAR cycle 15 observations (project LC15\_003) which have poor sensitivity as they were deliberately observed at low elevation due to a different science aim.

The data were processed following the procedure described by \cite{Tasse_2021}. This approach was also used for the previous public data release of the LoTSS Deep fields (\citealt{Tasse_2021}, \citealt{Sabater_2021}, \citealt{Kondapally_2021}, \citealt{Duncan_2021} and \citealt{Best_2023}) and is used for the wider area LoTSS (e.g. LoTSS-DR2; \citealt{Shimwell_2022}). The data are calibrated and imaged accounting for effects that are assumed constant throughout the observation as well as those that vary with time and/or position across the field of view (primarily the ionosphere and imperfections in the beam model). For completeness, a brief summary of the processing is given below.

After the observations were conducted the data were flagged, averaged and for more recent observations (those after Sept 2018) they were also compressed by the observatory from 32 bits per float to 12 bits per float (saving approximately a factor of 4 in storage volume with negligible impact on data quality for our purpose). These operations were done using the software packages DP3 (\citealt{Dijkema_2023}), AOFlagger (\citealt{Offringa_2012}) and Dysco (\citealt{Offringa_2016}). The data are then sent to the LOFAR Long Term Archive (LTA)\footnote{\url{https://lta.lofar.eu/}} with these particular datasets being stored either at the Forschungszentrum J\"{u}lich\footnote{\url{http://www.fz-juelich.de}} or the SURF\footnote{\url{https://www.surf.nl/}} site. These data are presently retrievable through the LTA web interface, except for data from LC2\_024 which are hosted elsewhere but are still available for download\footnote{\url{https://github.com/nudomarinero/AWS_elais-n1_public_data}}.

The LTA datasets are large (290\,TB) and we processed them on compute clusters local to the archives (JUWELS\footnote{\url{https://www.fz-juelich.de/en/ias/jsc/systems/supercomputers/juwels}} and Spider\footnote{\url{https://doc.spider.surfsara.nl/en/latest/Pages/about.html}}) to minimise data transport. Firstly the 3C295 calibrator observations corresponding to each target observation were processed through the PreFactor\footnote{\url{https://github.com/lofar-astron/prefactor}} calibrator pipeline (recently migrated to the LOFAR Initial Calibration pipeline; LINC\footnote{\url{https://linc.readthedocs.io/en/latest/}}). This derives time-independent corrections for the bandpass, polarisation alignment (offset between XX and YY phase) and clock offsets between different stations. The target data were then processed with the PreFactor/LINC target pipeline which applies the derived calibrator solutions in addition to a single ionospheric Faraday Rotation Measure correction across the field (\citealt{Mevius_2018}) before calibrating them against a sky model derived from the TIFR  GMRT  Sky  Survey alternative data release (TGSS-ADR1; \citealt{Intema_2017}). The target pipeline also performs additional flagging of data contaminated by radio frequency interference and in our case removes the international LOFAR stations (which are not used for this study) and averages the data to 8\,s time resolution and 0.0975 MHz spectral resolution (a factor of up to 64 averaging) before Dysco compressing to 16 bits per float (adding negligible additional noise in our case). These averaging parameters were chosen to minimise the data volume whilst keeping time and bandwidth smearing effects tolerable (see e.g. Sect. 3.1 of \citealt{Shimwell_2019}). The PreFactor/LINC calibrator and target pipelines are described in detail by \cite{deGasperin_2019}.

Prior to performing direction and time-dependent calibration and imaging of the target field data we first subtracted sources outside the region of interest using the dataset with direction-independent calibration only. This decreases contamination from the sidelobes of these sources and allows us to focus on a smaller region of the sky in subsequent calibration and imaging steps, which lowers the computational cost. Furthermore, a tighter tessellation of facets (corresponding to a smaller patch of ionospheric and beam variations) for which we derive direction-dependent calibration solutions can be used without significantly increasing the number of calibration directions (and degrees of freedom) which helps maintain robust calibration and mitigates against the absorption of unmodelled emission. A similar approach is typically applied to LOFAR HBA low-declination fields where the N-S primary beam extension is very large (e.g. \citealt{Hale_2019}). To choose an appropriate region for this source subtraction we tested subtracting sources outside of square regions ranging from $4.2^\circ\times 4.2^\circ$ to $7.0^\circ\times 7.0^\circ$ for target observation 798146 (selected as it is one of the best quality observations; see \href{https://doi.org/10.5281/zenodo.14603969}{Tab. S1}). For each test region size a low resolution (around 45$\arcsec$) $25^\circ\times 25^\circ$ image of the observation was created and sources outside the desired region were subtracted from the $uv$-data using the clean component models derived during the imaging. The subtracted data were then calibrated and imaged using DDF-pipeline\footnote{\url{https://github.com/mhardcastle/ddf-pipeline}} which corrects for ionospheric distortions and other direction-dependent or independent errors in the data. The pipeline uses kMS (\citealt{Tasse_2014} and \citealt{Smirnov_2015}) for simultaneous direction-dependent calibration and DDFacet (\citealt{Tasse_2018}) for imaging with these solutions applied. All DDF-pipeline runs were conducted with 45 directions and between 10,000$\times$10,000 and 17,000$\times$17,000 1.5$\arcsec$ pixels to encompass the area where sources have not been subtracted. The quality of the resulting images was assessed by comparing the number of sources detected within 1.5$^\circ$ of the pointing centre. Here we found that when the region remaining after subtraction was too small (in this case less than $5.0^\circ\times 5.0^\circ$) the calibration was more susceptible to diverging in some directions (those that contained the least flux density) and the source density on the images was on average lower. For the tests with regions larger than this the image source density was comparable (between 7500 and 7750) with the $5.83^\circ\times 5.83^\circ$ box providing the largest number of sources. 

The direction-independent subtraction of sources outside the central $5.83^\circ\times 5.83^\circ$ box was repeated for all epochs. The sky model derived from the target observation 798146 DDF-pipeline run was then used to derive direction and time-dependent calibration solutions in 45 directions for the other 71 target datasets. This calibration was done by providing DDF-pipeline with an input sky model and facet layout. This approach speeds up the processing significantly as it reduces the number of calibration cycles that is done from 8 to 4 and the number of imaging cycles from 7 to 2 (see Algorithm 1. of \citealt{Tasse_2021}). In this mode the pipeline only performs the following steps: fast direction-dependent and direction-independent calibration from the provided sky model; imaging with calibration solutions to derive an updated sky model; fast and slow direction-dependent calibration using the updated sky model; and final imaging with both fast and slow solutions applied. The computational cost for the various processing steps for all the data on an AMD Rome node with 64 
cores was approximately as follows: 
Prefactor/LINC 300\,hrs; direction-independent source subtraction 1000\,hrs; direction-dependent and 
independent calibration off a sky model 1500\,hrs; imaging to create an updated sky model 380\,hrs; final direction-dependent fast and slow calibration 1400\,hrs; final imaging 120\,hrs. We note that the imaging steps required processing all the data together but calibration steps were run simultaneously either for individual datasets (LINC/PreFactor and source subtraction) or portions of them (after LINC/PreFactor the data are stored in 1.95\,MHz blocks which are independently calibrated by kMS). Assuming that 20 nodes are used for all steps that can run simultaneously and that 1 node 
is used for imaging the total runtime is approximately 1 month.

\begin{figure}[htbp]
   \centering
   \includegraphics[width=\linewidth]{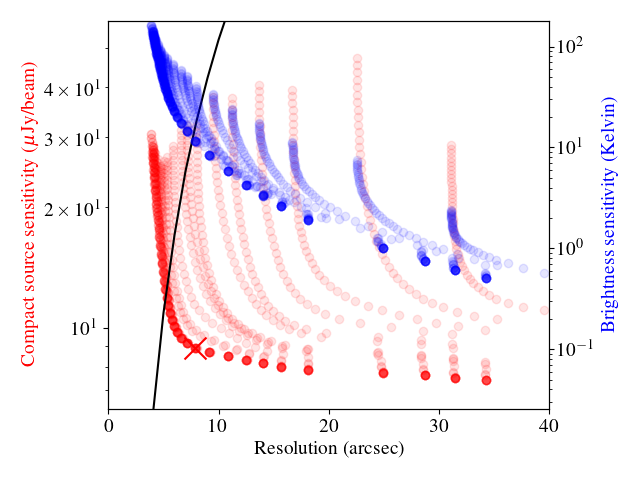}
   \vspace{-0.7cm}
   \caption{The expected compact source sensitivity (red) and brightness sensitivity (blue) as a function of imaging resolution for the 122-124\,MHz data from observation 798146. The resolution is varied by altering the visibility weightings with robust and tapering settings. The black curve shows the anticipated confusion noise (defined as 1 source per 10 resolution elements). The red cross shows the location of the 6$\arcsec$ resolution imaging parameters we have used (robust=-0.5, no taper). Only a small dataset was used for this analysis to limit the computational cost. }
   \label{Fig:vary_res}
\end{figure}

Prior to creating the final images we explored a range of different imaging parameters with the intention of creating images with optimal point-source sensitivity and surface brightness sensitivity. To find the appropriate settings we created 1000 different images spanning a wide range of robust parameters (-2.0 to 0.0 in intervals of 0.05) and outer $uv$-plane tapers (0 to 50,000$\lambda$ with intervals of 2000$\lambda$) using 122-124\,MHz data from observation 798146; the full range of robust parameters was not explored because the DDFacet fitting of the synthesized beam is somewhat unstable at robust parameters above 0 due to the highly non-Gaussian shape of the synthesized beam. In order to test the imaging parameters, we only produced Stokes V images allowing us to study just the thermal noise. This removed the need for deconvolution, eliminated any influence from confusion noise, and mitigated the dynamic range limitations associated with bright Stokes I sources. For each of the 1000 images we calculated the sensitivity (root mean square noise, $\sigma_{RMS}$) as well as the brightness temperature sensitivity ($\frac{\lambda^2}{2k_B\Omega_{bm}}\sigma_{RMS}$ where $\lambda$ is the wavelength, $\Omega_{bm}$ is the synthesized beam solid angle and $k_B$ is the Boltzmann constant). The measured sensitivities as a function of resolution (derived by DDFacet fitting the Point Spread Function, PSF, with a Gaussian) are shown in Fig. \ref{Fig:vary_res} where the values are scaled to expectations from the full 505\,hr ELAIS-N1 dataset. We also show the expected level of source confusion, which we define as the anticipated  flux density level at which there is one source per 10 resolution elements (see Sect. \ref{sec:confusion} for details on the confusion noise). The parameters we have chosen for imaging (marked on the figure) at 6$\arcsec$ resolution are a compromise between resolution and sensitivity.

\begin{figure*}[htbp]
   \centering
   \includegraphics[width=\linewidth]{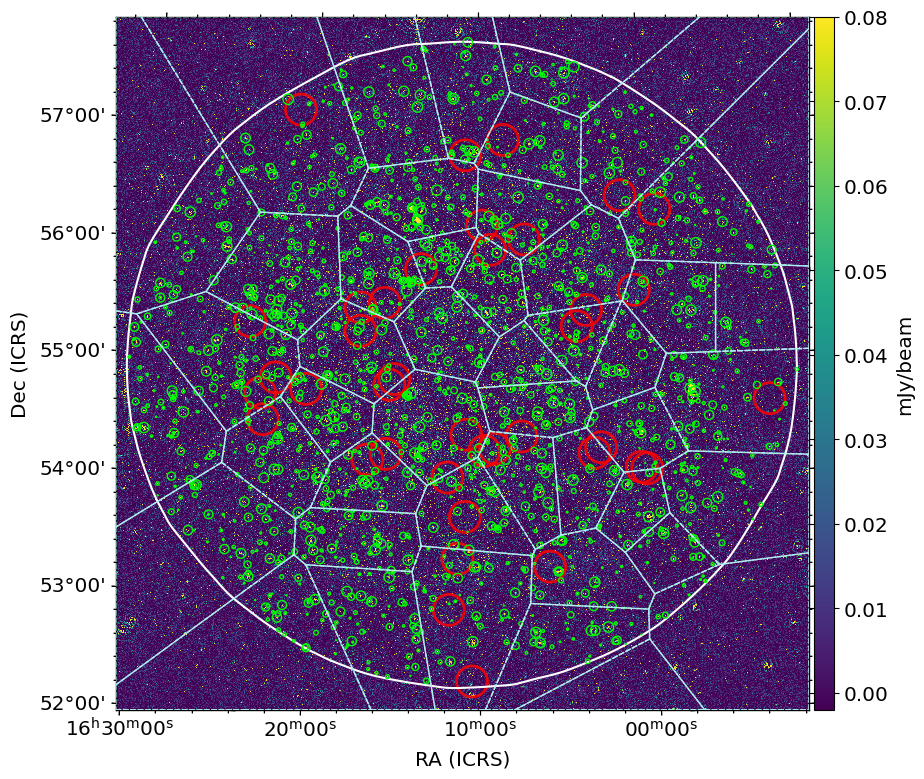}
    \includegraphics[width=0.45\linewidth]{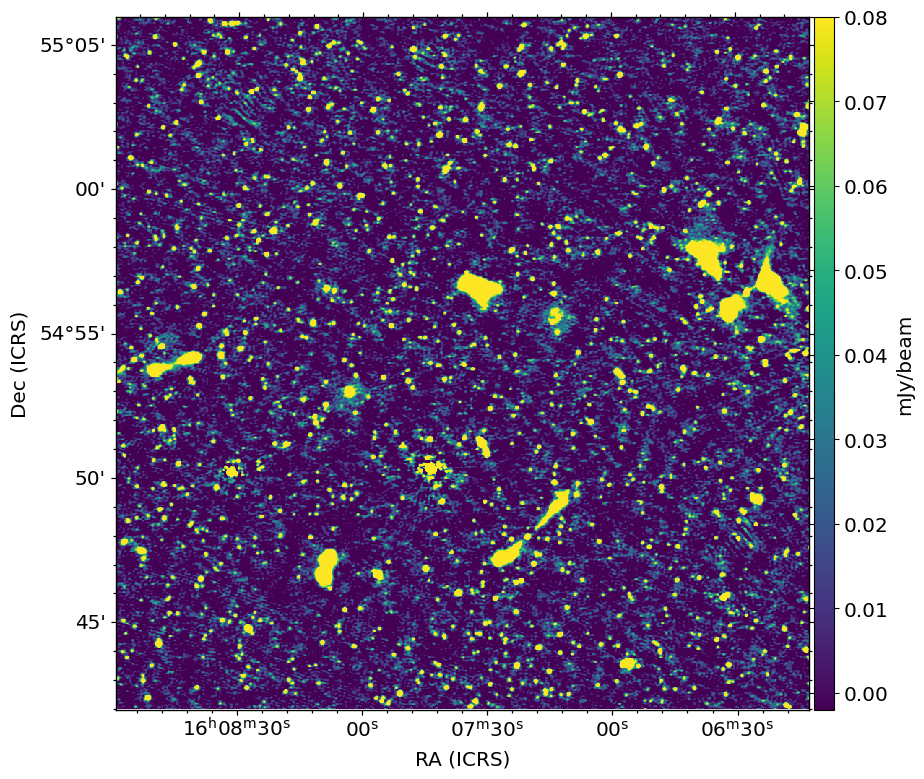}
    \includegraphics[width=0.45\linewidth]{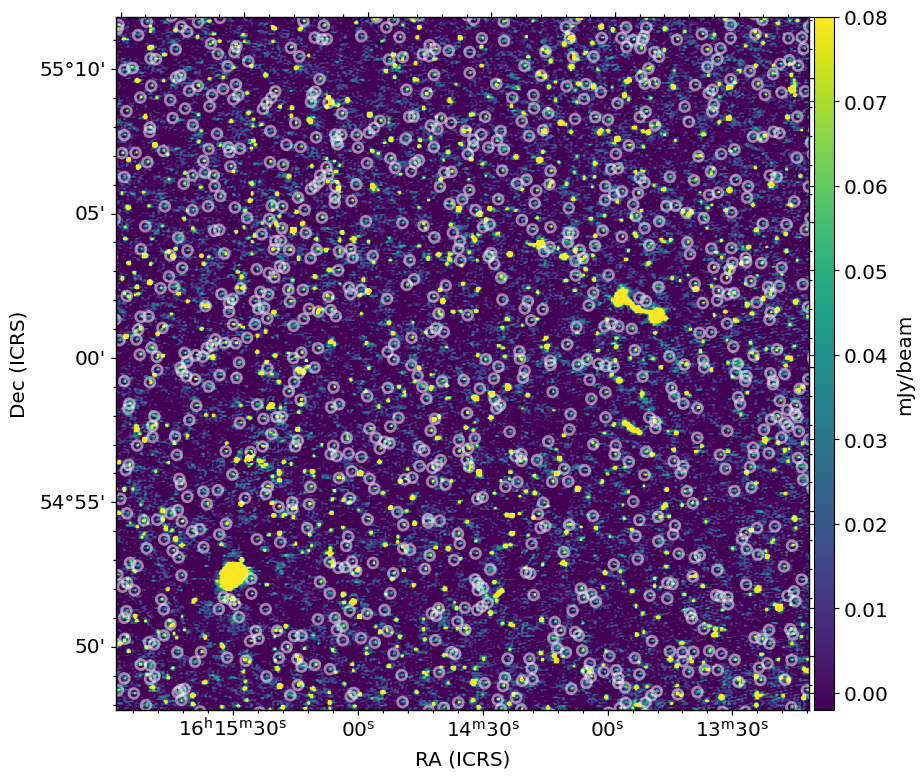}
    \vspace{-0.2cm}
   \caption{The top panel shows the full depth (10.7\,$\mu$Jy beam$^{-1}$), full area ($5.83^\circ\times 5.83^\circ$) $6\arcsec$ ELAIS-N1 LoTSS Deep image. The 30\% level of the power primary beam is shown in white (approximately a circle of radius 2.8$^\circ$) and encompasses 24.53 square degrees. The boundaries of the 45 facets used for direction-dependent calibration are shown with light blue lines. The green circles show regions where the noise is limited by the dynamic range (see Sect. \ref{sec:dynrange}) and the larger red circles show the locations of the most variable sources identified in the field (see Sect. \ref{Sec:variablity}). The  bottom panels show two example $0.4^\circ\times 0.4^\circ$ regions in more detail to highlight the source density (over 10,000 per square degree) and the variation in morphology of detected sources. The grey markers on the bottom right panel show sources in the new ELAIS-N1 catalogue that were not detected by \cite{Sabater_2021}.}
   \label{Fig:ELAIS_image}
\end{figure*}

The final 6$\arcsec$ resolution Stokes I image is shown in Fig. \ref{Fig:ELAIS_image}. The sensitivity at the field centre of the image is 10.7\,$\mu$Jy beam$^{-1}$ where approximately half of this is due to confusion noise -- see Sect. \ref{sec:confusion}.

\begin{table}
\renewcommand{\arraystretch}{1.15} 
\caption{A summary of ELAIS-N1 LoTSS Deep observations as well as the final image and calibrated data.}
 \centering
 \label{LOFAR-obs}
\begin{tabular}{|c|c|}
\hline
Coordinates (J2000) & 16:11:00 55:00:00 \\ \hline
Calibrated data frequency & 0.0975 \\  resolution (MHz) &  \\\hline
Calibrated data frequency & 115-177 \\  range (MHz) &   \\\hline
Calibrated data (calibration & 4.7 (0.1) \\  solutions) volume (TB)    & \\ \hline
Calibrated data time resolution (s) & 8 \\\hline
On-source integration time (hrs) & 505 \\\hline 
Calibrated data flagged fraction & 0.17 \\ \hline
Image thermal noise ($\mu$Jy beam$^{-1}$) & 7.5 at 6$\arcsec$ resolution  \\ \hline
Image central RMS ($\mu$Jy beam$^{-1}$)  & 10.7 at 6$\arcsec$ resolution \\ \hline
 \end{tabular}
\end{table}

\subsection{Source cataloguing}
\label{Sec:Sourcefinding}
We use \textsc{PyBDSF} (\citealt{Mohan_2015}; version 1.10.3) for the detection and characterisation of sources within the image.  \textsc{PyBDSF} was also used in the previous LoTSS Deep data release \citep{Sabater_2021, Tasse_2021} and the wider area LOFAR surveys \citep{Shimwell_2019, Shimwell_2022,deGasperin_2023}. However, the ELAIS-N1 LoTSS Deep image from this release has an especially high density of sources and is severely confusion noise limited (e.g. Fig. \ref{Fig:ELAIS_image}). To account for this we refine the set
up of \textsc{PyBDSF} compared to previous studies in order to accurately describe the noise in the image, optimise the number of sources detected and improve the characterisation of complex emission.

In our previous setup \textsc{PyBDSF} calculated an RMS map 
using a 2D sliding box across the image. This map was then used to identify (and fit) islands of contiguous emission above a threshold level.
With this method, sources are not removed from the image prior to the calculation of the RMS map. 
Therefore, for images with a high source density the RMS map is elevated compared to if sources were removed from the image prior to the calculation of the RMS map. This will affect the detection threshold across the image and, as such, will reduce the completeness of the resulting source catalogue. To mitigate this, we developed a multi-step \textsc{PyBDSF} process where we first derive an RMS map that better reflects the image noise and then supply this RMS map to \textsc{PyBDSF} to generate the catalogue for this work. We detail this process in Appendix \ref{appendix_pybdsf}.

Our final catalogue contains 154,952 sources across the image formed from 182,184 Gaussian components. For over 50\% of these sources this is their first detection at radio wavelengths. The catalogue spans 24.53 square degrees, within the 30\% level of the power primary beam, and extends down to a peak detection significance threshold of 4$\sigma_{thresh}$ and an island boundary threshold of 3$\sigma_{isl}$ (using the \textsc{PyBDSF} parameters  \texttt{thresh\_pix}  and \texttt{thresh\_isl} respectively).

\subsection{Source detection threshold}
\label{sec:thresholding}

In addition to refining the source cataloguing procedure we also revisited the detection threshold. In previous LoTSS and LoTSS Deep data releases we used a 5$\sigma_{thresh}$ peak threshold for cataloguing. Given that in LoTSS Deep fields we are able to match a high fraction of radio sources with optical counterparts (achieving > 97\% identification rate in LoTSS Deep; see \citealt{Kondapally_2021}), we can ascertain whether radio detections are likely to be genuine at a given detection threshold. We therefore investigated how varying our detection threshold influenced the number of spurious sources introduced.

Source extraction down to 4$\sigma_{thresh}$ using the refined \textsc{PyBDSF} procedure results in $\approx 30\%$ more sources than using the old procedure with a 5$\sigma_{thresh}$ threshold.  The increase in numbers is due to both the lower detection threshold and a higher completeness achieved at $5 < $SNR$ \leq 6$ due to the deeper RMS map (where we define SNR as the catalogued integrated flux density, $S_I$, divided by the integrated flux density uncertainty, $\sigma_{S_I}$). If we were to use a 3$\sigma_{thresh}$ threshold this would increase the source number by a further $\approx 20\%$. To determine whether it is appropriate to use a threshold below 5$\sigma_{thresh}$ we examined the optical counterparts of the radio sources allowing us to ascertain which radio detections are likely genuine. To do this we first created a very deep radio catalogue by repeating the cataloguing procedure in Sect. \ref{Sec:Sourcefinding} but using a lower detection limit of {\tt thresh\_isl=2.0} and {\tt thresh\_pix=3.0}.
We then considered only radio sources with major axis sizes $<$ 10$\arcsec$ from the catalogue (25\% of $3 < SNR \leq 4$ sources, 19\% of $4 < SNR \leq 5$ sources and 16\% of $5 < SNR \leq 6$ sources), as these are more suitable for the likelihood ratio (LR) cross-matching analysis which determines the ratio of the probability that a optical source is associated with a particular radio source, to the probability that it is unrelated \citep[e.g.][]{Saunders_1992,Smith_2011}. We used the multi-wavelength catalogue and the LR method, both as described by \citet{Kondapally_2021}, to determine the multi-wavelength cross-matches for these radio sources.

\begin{figure}
    \centering
    \includegraphics[width=\columnwidth]{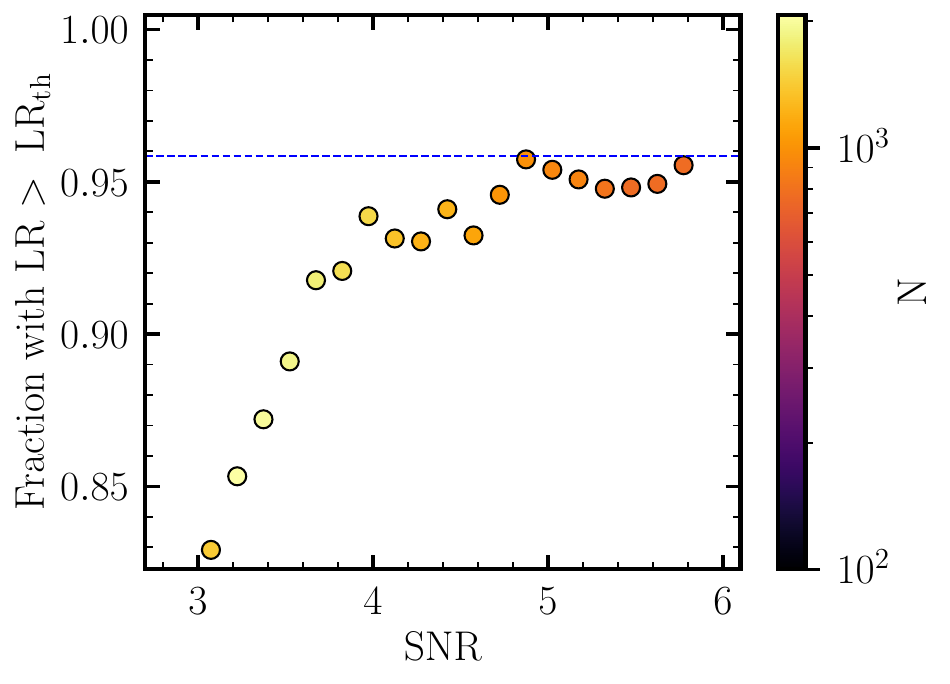}
    \vspace{-0.7cm}
    \caption{The fraction of small (major axis $<$ 10$\arcsec$) radio sources with a likelihood ratio match as a function of the SNR ($S_I/\sigma_{S_I}$). The colour of the points corresponds to the number of sources in each SNR bin. The blue horizontal line corresponds to the overall LR identification rate obtained. We find that lowering the detection threshold to a peak detection significance of 4$\sigma_{thresh}$ results in a large fraction ($\sim30\%$) of additional, genuine radio sources compared to adopting a 5$\sigma_{thresh}$ threshold. However, this also leads to a small increase in false detections (by $\sim$ 2-3\%).}
    \label{fig:lr_vs_snr}
\end{figure}

In Fig.~\ref{fig:lr_vs_snr}, we plot the fraction of sources that have a robust cross-match (defined based on the LR value that achieves a completeness and reliability of $>99\%$; see \citealt{Kondapally_2021}) as a function of SNR. The colour bar corresponds to the number of sources in each bin. The blue horizontal line corresponds to the overall cross-match rate for the full radio sample of $\sim$ 96 per cent; this is less than 100 per cent as the host galaxies for a small fraction of the sources will be too faint to be detected and hence missing from the multi-wavelength catalogue, while some radio sources may form part of a larger, more complex multi-component source which would not be expected to have a cross-match based on the use of the LR method alone. We find that at high SNRs the cross-match rate is close to the blue line, and then decreases below SNR $\sim$ 5; this is caused by some of the additional lower significance sources not being genuine radio sources. Based on this analysis we find that at $4 < SNR \leq 5$, around 2-3 per cent of the radio sources are likely not genuine. As the SNR decreases further down to SNR  $\sim$ 3, the fraction of matches declines more rapidly, and we find that a considerable fraction ($\sim$ 10 per cent) of the additional sources detected are likely not genuine, which would add significant contamination to the catalogue. We therefore adopt the use of a 4$\sigma_{thresh}$ detection threshold for source extraction. This allows us to catalogue 30\% more sources than our previous 5$\sigma_{thresh}$ approach and the vast majority of these additional sources we believe are genuine detections of faint radio sources. We do expect a small increase in the fraction of non-genuine sources by detecting down to 4$\sigma_{thresh}$ compared to 5$\sigma_{thresh}$ and therefore urge caution when conducting studies that include these faint sources. We also note that our upcoming efforts for visual classification and inspection will help identify the spurious sources \citep[see][]{Williams_2021,Kondapally_2021}.

\section{Image quality}
\label{Sec:image_quality}

The processing strategy used for this data release follows the same approach used for the previous ELAIS-N1 data release in \cite{Sabater_2021}, with the exception of the additional source subtraction we performed to remove sources outside the field of view. However, the amount of data used is approximately three times larger (505\,hrs compared to 163\,hrs) and due to ionospheric activity and telescope performance we expect substantial variations in quality between different observations. Below we assess several key image quality properties and, where possible, quantify appropriate uncertainties on these for this data release.

\subsection{Source extensions}
\label{sec:source_extensions}

Slightly extended sources and point-like sources cannot be definitively distinguished in our images due to both measurement errors and imperfections in the image quality. For example, for a source to have the exact shape of the restoring beam it must be represented by a single pixel in our clean component model image. However, even for isolated seemingly compact sources we find that the flux density on the clean component model image is generally distributed between many neighbouring pixels. Typically, high SNR sources only have about 20-50\% of their integrated flux density at the position of the peak pixel whereas low SNR sources generally have a higher fraction of their integrated flux density in a single pixel. This is likely a consequence of only deconvolving to a certain depth. Furthermore,  in our images the blurring of sources noticeably increases as a function of distance from the field centre. This is primarily due to time and bandwidth smearing but also the outer facets are larger and thus have greater ionospheric variations within the facet leading to poorer calibration. During the self-calibration this phase blurring is difficult to rectify because it is present in the sky models being used for the calibration.

Thus, to establish a criterion to distinguish unresolved from extended sources we must account for all sources being inherently blurred. To do this we first select and characterise a sample of 45,676 isolated seemingly compact sources from the full catalogue of 154,953 sources. We define these sources as those that are more than 15$\arcsec$ from their nearest neighbour, have \textsc{PyBDSF} code `S' corresponding to being fit by a single Gaussian, a major axis below 15$\arcsec$, lie in the deconvolution mask and within 30\% of the power primary beam and have $\frac{S_I}{\sigma_{S_I}} > 5$. Firstly we plotted the ratio of the integrated flux density, $S_I$, to peak brightness ($S_P$) versus distance from the pointing centre for this population and found the median $\frac{S_I}{{S_P}}$ in different radial bins. From these median values we derived a best fit of  $\frac{S_I}{{S_P}}=1.038\times e^{0.094\times D}$ where $D$ is the distance from the field centre in degrees. We then modified all ${S_P}$ values using this function (i.e. multiplying by $1.038\times e^{0.094\times D}$) which on average corrects for the offset in $\frac{S_I}{{S_P}}$ from unity at the field centre as well as the radial dependence, thus giving a more constant $\frac{S_I}{{S_P}}$ for all sources. Finally, we attempted to define a distinction between extended and unresolved sources using our modified catalogue and the behaviour of $\frac{S_I}{{S_P}}$ (a proxy for source extension) versus $\frac{S_I}{\sigma_{S_I}}$.

For our defined population of seemingly compact sources in our modified catalogue, the distribution of $\frac{S_I}{{S_P}}$ at a given $\frac{S_I}{\sigma_{S_I}}$ is skewed with an excess of sources at high $\frac{S_I}{{S_P}}$ values. Furthermore, as $\frac{S_I}{\sigma_{S_I}}$ increases the number of sources significantly decreases which makes it challenging to fit the distribution of $\frac{S_I}{{S_P}}$. Both these aspects are expected even for a population of perfect compact sources with the skewness arising from the existence of a correlation between uncertainties, i.e. if, due to measurement errors, the recorded ${S_I}$ is higher than the real source ${S_I}$ then the ${S_P}$ will also likely be higher than its true value. To somewhat mitigate these issues and attempt to characterise the behaviour of seemingly compact sources in our images we conducted a suite of simulations.

We generated mock images by injecting a population of sources with flux densities taken from a given distribution of integrated flux densities and morphologies derived from  our real \textsc{clean} component models (from our DDFacet model images) of the seemingly compact sources identified using the criteria above. To do this the \textsc{clean} component models were injected into a blank image and scaled using the distribution of integrated flux densities taken from the \cite{Mandal_2021} counts above 0.43\,mJy beam$^{-1}$ and extended to lower flux density (1\,$\mu$Jy beam$^{-1}$) using the number counts from the T-RECS (\citealt{Bonaldi_2023}) simulations (see Sect. \ref{sec:confusion}). The \textsc{clean} component model used for a particular injection was randomly selected from all the available models of sources that are close in integrated flux density to the source we are injecting. We then scaled the \textsc{clean} component models to precisely match the intended integrated flux density. The simulated image containing all the deconvolution models was then convolved with the restoring beam (6$\arcsec$) and added to the \textsc{PyBDSF} residual image to produce a mock image which we then catalogued using \textsc{PyBDSF} following the procedure in Sect. \ref{Sec:Sourcefinding}. To reproduce the radial effects observed in the real images we inject sources at the same radial distance as they had in the real observations. We then again derive a function that describes the offset and radial dependence in $\frac{S_I}{{S_P}}$ for the simulated image and modify the catalogued $S_P$ values to on average correct for this. 

\begin{figure}[htbp]
   \centering
   \includegraphics[width=\linewidth]{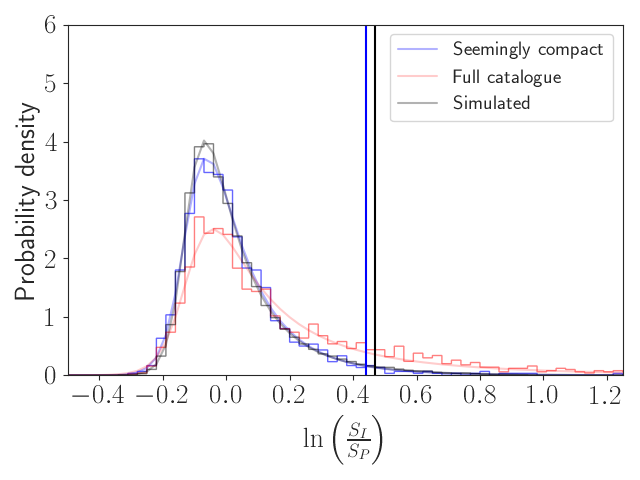}
   \includegraphics[width=\linewidth]{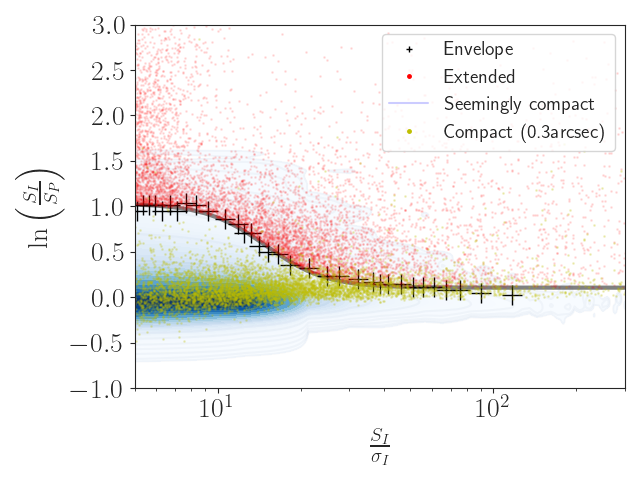}
   \vspace{-0.7cm}
   \caption{Top: Histograms showing the probability density (bin width  0.03) of $\ln(\frac{S_I}{{S_P}})$ for sources with $15.8 < \frac{S_I}{\sigma_{S_I}} < 17.3$. This corresponds to 999 seemingly compact LoTSS sources (blue), 1673 LoTSS sources in the full catalogue (red) and 35286 sources in our simulations of seemingly compact sources (grey) described in Sect. \ref{sec:source_extensions}. The fits to the histograms use a normal-inverse Gaussian distribution and are shown with  faint solid lines. The vertical lines show the cutoffs we have chosen to separate unresolved from extended sources, these are defined as the location where the complement of the Cumulative Distribution Function for the full catalogue is equal to 5 times that of the simulated seemingly compact (black) or seemingly compact source (blue) distributions. Hence sources with $\ln(\frac{S_I}{{S_P}})$ above this value are 5 times more likely to lie in the tail of the red curve compared to the black or blue curves. We use this definition as a boundary to separate unresolved from extended sources. Bottom: The separation values for unresolved and extended sources derived from the simulations as a function of $\frac{S_I}{\sigma_{S_I}}$ are shown with black crosses which have been fit with a sigmoid function (grey). The blue density contours and colour scale show the seemingly compact LoTSS source population whereas the red points show the locations of the sources in the full LoTSS catalogue that are extended according to the criteria we have used. The yellow points show the locations of sources in our catalogue that are also detected by \cite{deJong_2024} at 0.3$\arcsec$ resolution.}
   \label{Fig:source_sizes}
\end{figure}

Repeating this procedure a number of times (we performed 20 runs) provides us with a large number of sources and allows us to better constrain the distribution of $\frac{S_I}{{S_P}}$ for seemingly compact sources at a given $\frac{S_I}{\sigma_{S_I}}$. An example histogram showing the distributions of simulated sources together with the seemingly compact sample and the full LoTSS sample for a moderate $\frac{S_I}{\sigma_{S_I}}$ is shown in Fig. \ref{Fig:source_sizes}. It is clear from this example that separating unresolved and extended sources cannot be done definitively. We chose to separate sources where the survival fraction (the complement of the Cumulative Distribution Function) of the simulated seemingly compact sources was a factor of 5 times lower than the survival fraction of the full LoTSS catalogue (as shown in the Figure, comparable values are found if we use the real seemingly compact source sample instead of the simulated one). This implies that sources with $\frac{S_I}{{S_P}}$ equal to or exceeding this value are five times more likely to be associated with the tail of the full LoTSS catalogue $\frac{S_I}{{S_P}}$ rather than the simulated seemingly compact one. This is shown as a function of $\frac{S_I}{\sigma_{S_I}}$ in the bottom panel of Fig. \ref{Fig:source_sizes} and the best fit sigmoid function to this is:
\begin{equation}\label{eq:resolved_sources}
R_{5} = 0.10 + \left( \frac{0.93}{1+\left( \frac{S_I/\sigma_{S_I}}{14.95}\right)^{3.99}} \right) 
\end{equation}
Using these constraints we find that for the full LoTSS Deep catalogue where a radial dependence correction has been applied (see above) a total of 14,025 (9\%) out of 154,952 sources are extended with the rest being unresolved at our resolution.

We further validate our unresolved criteria through comparison with the 0.3$\arcsec$ resolution \cite{deJong_2024}  catalogue. At this high resolution the surface brightness sensitivity is low (see e.g. Fig. \ref{Fig:vary_res}) but the sensitivity to compact sources, at 14\,$\mu$Jy beam$^{-1}$, is only 30\% higher than in our catalogue. We therefore expect \cite{deJong_2024} to be primarily sensitive to compact sources and this is demonstrated in Fig. \ref{Fig:source_sizes} (see also e.g. \citealt{Sweijen_2022}). Here only 11\% of sources crossmatched (within 1$\arcsec$) between the full \cite{deJong_2024} catalogue and our full catalogue are deemed extended at 6$\arcsec$ by our criteria, where almost all of these possibly extended sources have high $S_{I}/\sigma_{S_I}$. If we only examine the fainter sources in our catalogue ($S_{I}/\sigma_{S_I} < 15$) that are detected by \cite{deJong_2024} then just 0.3\% of these are classified as extended at 6$\arcsec$ resolution by our criteria. This gives confidence in our criteria to separate unresolved from extended sources. The comparison also highlights that large numbers of the sources we define as unresolved are highly resolved by LOFAR when utilising the full ILT 0.3$\arcsec$ resolution. For example, only 22\% of sources with $S_I/\sigma_{S_I} > 15$ in our catalogue are detected at 0.3$\arcsec$ by \cite{deJong_2024} with the rest being undetected likely due to surface brightness sensitivity limits. If we remove sources classified as extended at 6$\arcsec$ resolution this increases to  35\% indicating that our unresolved criteria only slightly helps to select sources that are still compact at 0.3$\arcsec$.

\subsection{Flux density scaling and uncertainty}
\label{fluxscale}

In \cite{Sabater_2021} the flux density scale of the ELAIS-N1 LoTSS Deep DR1 image was carefully examined using auxiliary measurements from VLSSr at 74MHz (\citealt{Lane_2014}, TGSS-ADR at 150\,MHz (\citealt{Intema_2017}), 6C at 151\,MHz (\citealt{Hales_1990}), the ELAIS-N1 GMRT survey at 325\,MHz (\citealt{Sirothia_2009}), WENSS at 325\,MHz (\citealt{Rengelink_1997}), the ELAIS-N1 GMRT survey at 610\,MHz (\citealt{Garn_2008}), NVSS at 1.4\,GHz (\citealt{Condon_1998}) and FIRST at 1.4\,GHz (\citealt{Becker_1995}). The procedure used by \cite{Sabater_2021} and the large amount of auxiliary data in the region allowed for them to achieve an uncertainty of the flux density scale of their LOFAR maps of 6.5\% in the central region. This is better than typically obtained with the LOFAR HBA where the accuracy is generally limited to 10-20\%. This limitation is thought to be partly a consequence of inaccuracies in the LOFAR beam model as it is found that transferring calibration solutions derived from calibrators (whose models are on the \citealt{Scaife_2012} flux density scale) does not necessarily result in an accurate flux density scale of the target field (e.g. \citealt{Hardcastle_2016} and \citealt{Shimwell_2022}).

To again obtain a high accuracy in the flux density scale we simply align our flux density scale with the catalogue of \cite{Sabater_2021}. To do this, from our catalogue we first select isolated (those without a nearest neighbour within 15$\arcsec$), unresolved (defined in Sect. \ref{sec:source_extensions}) sources that have \textsc{PyBDSF} code `S'. 
This leaves  97,678 sources (which we refer to throughout as the unresolved, isolated sources in our catalogue) to cross-match with the \cite{Sabater_2021} catalogue. Of these, two sources match (within 5$\arcsec$) multiple sources in the \cite{Sabater_2021} catalogue and are discarded. If we then do a further cut to remove sources not detected in the \cite{Sabater_2021} catalogue and also low significance sources in either catalogue ($S_I/\sigma_{S_I} < 10$) then this reduces the number of cross matched sources to 18,390. 
We find a median ratio of 0.89 between the integrated flux densities from \cite{Sabater_2021} and those from the new catalogue (differences on this scale are typical and reflect LOFAR HBA flux density scale calibration inaccuracies).  Our map and catalogue are scaled by this value in order to put them on the same flux density scale as \cite{Sabater_2021}. This flux density scale is also consistent with \cite{deJong_2024} and after removing sources detected in either image at $S_I/\sigma_{S_I} < 10$ the median of our integrated flux density measurements over those from \cite{deJong_2024} is 0.97. We note that due to time and bandwidth smearing as well as calibration errors there are larger variations in the peak brightness (see Sect. \ref{sec:source_extensions} and \cite{deJong_2024} Sect. 6.2).

For completeness, and because new auxiliary data are available, 
we again attempt to verify our flux density scale and further assess its accuracy. Instead of cross matching our catalogue with all possible radio wavelength catalogues covering the region, we only used the deeper, higher resolution catalogues for cross matching. These are the 54\,MHz LoLSS catalogue (see \citealt{deGasperin_2023} and \citealt{deJong_2024}), as well as the catalogues from the 325\,MHz (\citealt{Sirothia_2009}) and 610\,MHz (\citealt{Ocran_2020} and \citealt{Ishwara-Chandra_2020}) GMRT ELAIS-N1 surveys and the 1.4\,GHz FIRST survey (\citealt{Becker_1995}). The LoLSS catalogue used was not available at the time of the \cite{Sabater_2021} study and was created by \cite{deJong_2024} who processed a LoLSS dataset covering ELAIS-N1 using the standard data processing strategy for that survey (see \citealt{deGasperin_2023}). The flux density scales applied during the creation of the catalogued images that we compare with is consistent with the \cite{Scaife_2012} flux density scale, with the exception of the 325\,MHz catalogue. To correct the 325\,MHz catalogue we scaled it according to the 3C286 flux densities listed by \cite{Scaife_2012} compared to the values for this calibrator listed by \cite{Sirothia_2009}. We then created a cross matched catalogue using the appropriate cross matching radius given in Tab.3 of \cite{Sabater_2021} for comparison surveys in common. For LoLLS we used a 6$\arcsec$ cross matching radius and for the 610\,MHz GMRT survey we used 10$\arcsec$. Only 63 sources exist in the LoTSS Deep catalogue that are detected at all frequencies covered by the  auxiliary catalogues (if we were to remove sources based on the unresolved, isolated criteria given above we would be left with just 7 sources so we do not apply those cuts here).

\begin{figure}[htbp]
   \centering
   \includegraphics[width=\linewidth]{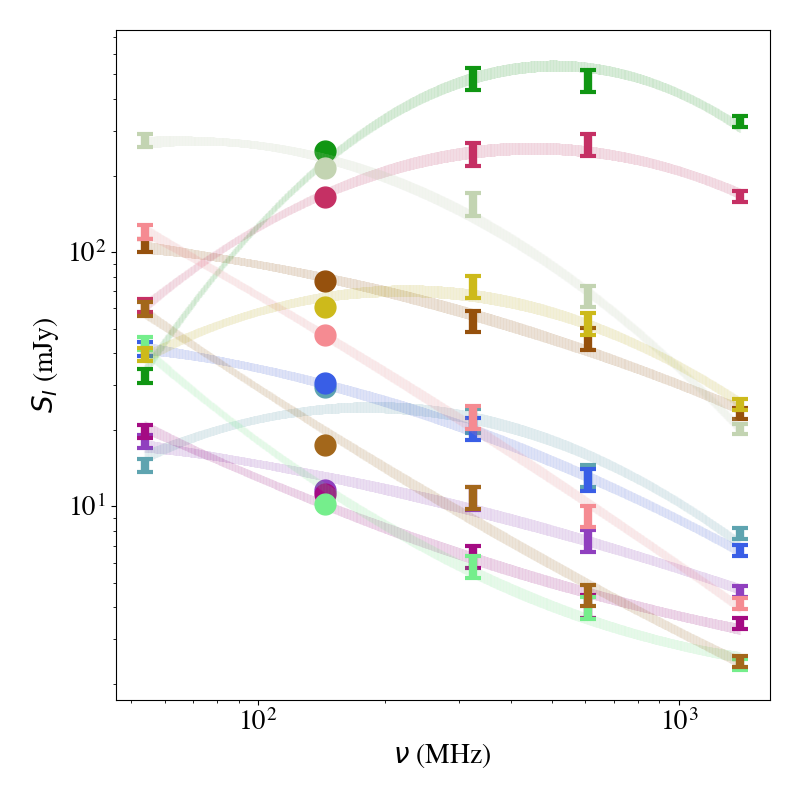}
   \vspace{-0.7cm}
   \caption{A selection of 12 of the 55 sources used to verify the deep ELAIS-N1 flux density, each shown in a different colour. The circles show the LoTSS Deep measurements and the error bars show the LoLSS, GMRT 325\,MHz, GMRT 610\,MHz and FIRST measurements with the associated error in the flux density scale. The lines show the derived second-order polynomial fits ($\log S_I(\nu) = \log S_0 + A \log(\nu/\nu_0) + B \log(\nu/\nu_0)^2$ where $S_0$, $A$ and $B$ are the best-fit parameters) with the line thickness showing the uncertainty in the fit.}
   \label{Fig:source_spectrum}
\end{figure}

For each of these sources we then fitted only the measurements in the auxiliary catalogues using a second-order polynomial to describe the log-log spectrum over frequency (see Fig. \ref{Fig:source_spectrum}) whilst taking into account the errors on the various flux density measurements by repeating the fitting a number of times but with flux densities drawn from a Gaussian distribution centred on the measured source flux density with a width reflecting the uncertainty. For these we assumed a 5\% flux density scale error for FIRST (\citealt{White_1997}), 6\% for LoLSS (\citealt{deGasperin_2023}) and 10\% for both the GMRT ELAIS-N1 surveys (\citealt{Chandra_2004}). During this process we found the spectrum of six sources was poorly fit by a 2nd order polynomial and excluded these from the remainder of the analysis. We also note that the \cite{Sirothia_2009} values were found to be systematically high by 25\%, so these were scaled down -- it is mentioned by \cite{Sirothia_2009} that their integrated flux densities were about 30\% higher than WENSS, which is consistent with the offset we find. The fitted polynomials were subtracted from the source measurements to give residuals for each fit at each frequency. From these residuals we found that the LoTSS Deep flux densities on average agree with the fitted polynomial giving confidence in the flux density scale but there is a standard deviation of 9\% which reflects an upper bound on the uncertainty. This derived uncertainty is comparable to what was found by \cite{Sabater_2021} for the previous LoTSS Deep data release. We note that our derived uncertainty is influenced by the uncertainty used for the comparison surveys, for example our uncertainty would increase to 11\% if we were to increase the errors on the comparison surveys to 10\%, 10\% and 15\% for FIRST, LoLSS and the GMRT surveys respectively.

\subsection{Astrometric accuracy}

During the data processing we refine our astrometry by aligning each individual facet used for calibration with Pan-STARRS DR2 (\citealt{Flewelling_2020}) following the procedure described in \cite{Shimwell_2019}. This results in a typical shift per facet of magnitude 0.44$\pm$0.20$\arcsec$ in RA and 0.44$\pm$0.16$\arcsec$ in Dec where these corrections are applied to the $uv$-data by DDFacet when constructing the final image.  To validate the astrometric precision of our final catalogues we again use Pan-STARRS DR2, which has a typical astrometric accuracy within 0.03$\arcsec$ (see e.g. \citealt{Magnier_2020} and \citealt{Makarov_2017}). To do this we construct a cross matched catalogue by performing a simple nearest neighbour cross-match between the unresolved, isolated sources in our catalogue and the mean Pan-STARRS DR2 epoch positions. In the cross matching we allow up to  1$\arcsec$ separation and only use Pan-STARRS DR2 sources that are detected in more than 5 single-epoch images to eliminate spurious detections.

\begin{figure}[htbp]
\includegraphics[width=\linewidth]{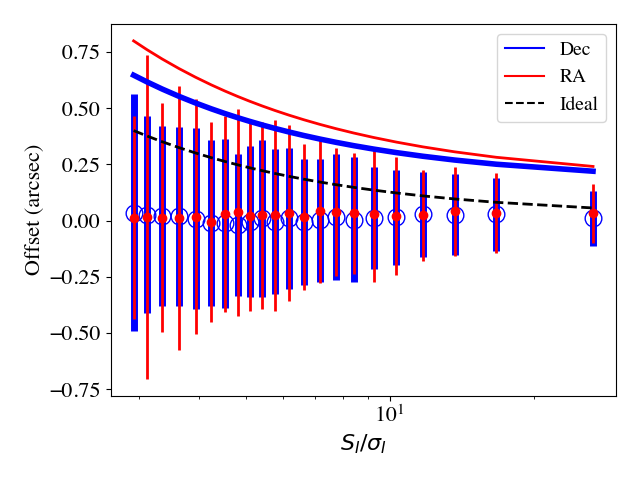}
    \vspace{-0.7cm}
    \caption{RA (red) and Dec (thick blue) astrometric offsets as a function of $S_{I}/\sigma_{S_I}$. Each flux density bin was chosen to contain 1000 sources and the histograms of the distribution of RA or Dec offsets between our catalogue and Pan-STARRS DR2 were fit with a Gaussian function. The points and error bars show the centre of the fit and $\pm$ the standard deviation respectively. Curves fitting the derived standard deviations of RA and Dec as a function of $S_{I}/\sigma_{S_I}$ are also shown with the same colour scheme.  The black dashed curve shows the behaviour observed through a simple simulation of perfectly compact sources injected into our residual image and characterised with \textsc{PyBDSF}.}
   \label{Fig:astrometry_SNR}
\end{figure}

Using all sources in this cross matched catalogue we first examined the offset between the optical and radio positions as a function of $S_{I}/\sigma_{S_I}$. Flux density bins were constructed such that they each contained 1000 sources and histograms of the RA and Dec offsets within each bin were fit with a Gaussian function. The centre of the Gaussian function indicates the accuracy of our alignment with Pan-STARRS DR2 whilst the standard deviation is scatter due to e.g. signal-to-noise effects and calibration imperfections. We found that for high significance ($S_{I}/\sigma_{S_I}>15$) detections the measured standard deviation is 0.2$\arcsec$ or less. For lower significance detections ($S_{I}/\sigma_{S_I}$ < 5), where signal-to-noise effects play a larger role, this increases to 0.5$\arcsec$. The width of this fitted Gaussian as a function of $S_{I}/\sigma_{S_I}$ can be described by $\sigma_{RA}$=$\sqrt{\left( \frac{1.83}{S_{I}/\sigma_{S_I}} \right)^2 + 0.17^2}$ and $\sigma_{DEC}$=$\sqrt{\left( \frac{1.40}{S_{I}/\sigma_{S_I}} \right)^2 + 0.17^2}$. The centre of the fitted Gaussian (i.e. the systematic positional offset) has a less clear dependence on $S_{I}/\sigma_{S_I}$ but for the previously defined high significance source sample it was found to be 0.04$\arcsec$ in RA and 0.03$\arcsec$ for Dec. For comparison with our measurements we performed a simple simulation where a population of perfect point-like sources were injected into the \textsc{PyBDSF} residual image and this simulated image was catalogued using \textsc{PyBDSF}. Here we again used the source counts derived from observations by  \cite{Mandal_2021} but extended to lower flux densities using the T-RECS simulations (\citealt{Bonaldi_2023}). In the simulations we found comparable curvature in the offsets as a function of $S_{I}/\sigma_{S_I}$ but far lower offsets at high $S_{I}/\sigma_{S_I}$. The results of the simulation together with the measurements from our images are shown in Fig. \ref{Fig:astrometry_SNR}.

\begin{figure}[htbp]
\includegraphics[width=\linewidth]{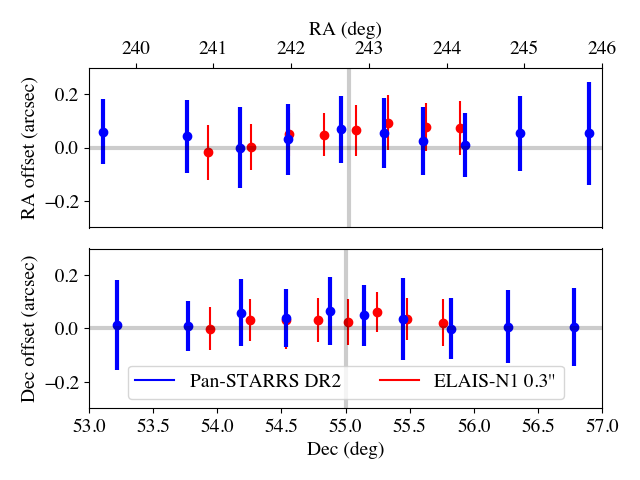}
\vspace{-0.7cm}
    \caption{RA and Dec offsets between our catalogued positions and Pan-STARRS DR2 (thick blue) and the ELAIS-N1 0.3$\arcsec$ image (red) from \cite{deJong_2024} as a function of RA and Dec respectively. The error bars show $\pm$ the standard deviation of a Gaussian fitted to histograms of the offsets and the points show the centre of the fitted Gaussian. The grey vertical line shows the field centre and the grey horizontal line shows 0 offset.}
   \label{Fig:astrometry}
\end{figure}

We also examined the offsets as a function of RA and Dec using the high significance source sample from the matched Pan-STARRS catalogue. The width of each RA or Dec bin was chosen so that each bin contained 200 sources. As shown in Fig. \ref{Fig:astrometry} the variations in the centres and standard deviations of fitted Gaussians are small although the outermost regions tend to have larger standard deviations. Finally, for completeness we also repeated this analysis but comparing with the 0.3$\arcsec$ resolution LOFAR image from \cite{deJong_2024}. In this analysis we used the same cross matching procedure as used for Pan-STARRS DR2 and for consistency applied a $S_{I}/\sigma_{S_I}>15$ cut to our catalogue and kept 200 sources per RA or Dec bin. There is again good astrometric alignment and from all sources in the cross matched catalogue we find the fitted Gaussian centres of RA and Dec offsets to be 0.05$\arcsec$ and 0.03$\arcsec$ respectively. The derived standard deviations of 0.11$\arcsec$ in RA and  and 0.10$\arcsec$ in Dec were even smaller than found through the comparison with Pan-STARRS DR2. The results from this cross matching are  shown in Fig. \ref{Fig:astrometry} and the variations are again quite small over the smaller field of view imaged by \cite{deJong_2024}.

We note that the LR catalogue from \cite{Kondapally_2021} was not used for astrometric verification because it does not cover the full LOFAR field of view. Furthermore, when it was constructed it made use of a preliminary Pan-STARRS data before the astrometry was refined by comparison with GAIA data (see \citealt{Magnier_2020}). As a result the multi-wavelength positions in \cite{Kondapally_2021} are on average 0.12$\arcsec$ in RA and 0.1$\arcsec$ in Dec offset from both Pan-STARRS DR2 positions and our new radio catalogue. Work is ongoing to construct an updated multi-wavelength catalogue with deeper optical datasets for identifications of the LoTSS Deep DR2 sources and this will have improved astrometry.

\subsection{Dynamic range}
\label{sec:dynrange}

To assess the dynamic range limitations of our final deep image we examine the noise enhancement within a 5$\arcmin$ radius around the 635 very isolated (more than 15$\arcsec$ from the nearest source of any flux density and more than 300$\arcsec$ from the nearest source with $S_I > 10$\,mJy) sources with apparent integrated flux densities exceeding $S_I > 5$\,mJy, where the region examined is fully within the 30\% of the power primary beam. For each source we measure the median absolute deviation (MAD) of the pixel values in annular regions of increasing distance from the source. We fit these profiles with a Gaussian function, remove sources where this is a poor fit and divide all the fitted Gaussian functions by the measured local RMS. We group the sources into 10 apparent integrated flux density bins each containing the same number of sources. At each radial bin we calculate the median of the MAD values derived from all the sources in the appropriate integrated flux density bin as well as the bootstrap errors. For each apparent integrated flux density bin we then fit these median MAD profiles with a Gaussian and find the point at which the median MAD is only 5\% higher than the background noise and calculate the area within this region. Finally we find that the area impacted by dynamic range versus integrated flux density is well described by an exponential function allowing us to derive that, for sources with $S_I$ greater than 5\,mJy:
\begin{equation}
    A_{Dyn,5\%} = -5.37\times10^{-4} + (6.62\times10^{-3} \times (1-e^{-22.7 S_I}))
\end{equation}
where $A_{Dyn,5\%}$ is the dynamic range limited area in square degrees and $S_I$ is the integrated flux density in Jy. We can use this to mask our image to contain only regions not substantially impacted ($<5\%$) by dynamic range.

We find that 7.4\% of the map within the 30\% of the power primary beam has noise levels that are at least 5\% higher than the local RMS due to dynamic range limitations, these regions are shown in Fig. \ref{Fig:ELAIS_image}.

\subsection{Individual observation imaging quality}
\label{Sec:indiv_quality}

To compare the individual image quality we first aligned all individual images to the same flux density scale by following the full procedure outlined in Sect.\,\ref{fluxscale} but using our deep catalogue for comparison rather than the \cite{Sabater_2021} catalogue. We then constructed a catalogue containing only isolated unresolved sources in the deep field catalogue that are detected in all 64 of the individual epoch images. 
To create this catalogue we further filtered our unresolved, isolated deep field catalogue so that it just contains $\frac{S_I}{\sigma_{S_I}}>5$ sources that are not in the 7.4\% of the image impacted by dynamic range limitations (Sect. \ref{sec:dynrange}).
We then cross-matched the individual epoch catalogues (unfiltered) with the filtered deep catalogue using a cross matching radius of 4$\arcsec$ and remove sources not present in all epochs. This yielded a catalogue of 560 sources that have integrated flux densities between 0.55 and 7.55\,mJy.

\begin{figure}[htbp]
   \centering
   \includegraphics[width=0.60\linewidth]{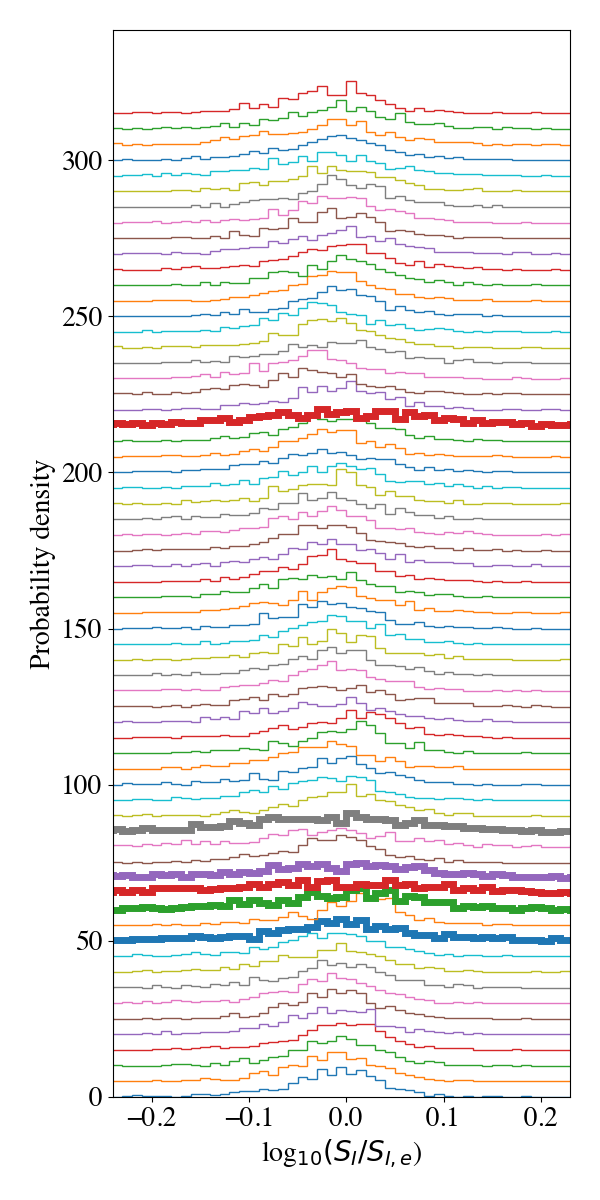}
   \includegraphics[width=0.9\linewidth]{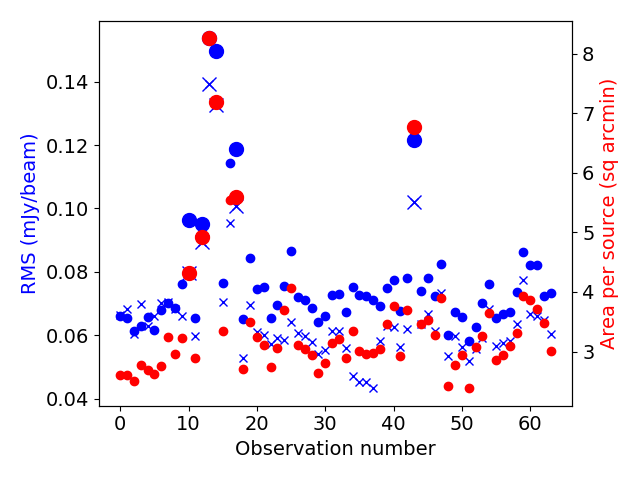}
   \vspace{-0.2cm}
      \caption{The top panel shows the probability density histograms of the integrated flux density ratio for the 64 individual epochs catalogues compared to measurements from our final continuum image. The $y$-axis of each epoch is offset by 5 for display purposes. The six thick lines correspond to observations L233804, L346136, L346154, L346454, L347494 and L782679 which have poor flux density values and were identified as having either a median flux density ratio or standard deviation further than 1.5 sigma from the other observations. In the bottom panel we show the RMS measured for each of the 64 epochs with blue circles and the blue crosses show the RMS scaled to a consistent 90$^\circ$ elevation, 231 subbands and 8hrs duration without any flagging (see \href{https://doi.org/10.5281/zenodo.14603969}{Tab. S1}). The red circles show the area per detected source in each of the epoch images (i.e. inverse of source density). The larger circles and crosses correspond to the six poor quality observations.}
   \label{Fig:epoch_flux_densities}
\end{figure}

The quality of the images derived from each observation varies substantially, as is apparent from \href{https://doi.org/10.5281/zenodo.14603969}{Tab. S1} which shows details of the individual epoch images and their corresponding observations. To further assess the relative quality of the images we examined the 560 sources that are detected in all 64 epochs. Histograms of the integrated flux density ratios of the sources in each epoch compared to the deep image are shown in Fig. \ref{Fig:epoch_flux_densities}. Whilst the median values of the flux density ratios (0.97$\pm$0.02) show the flux density scales of the images are well aligned the widths of the histograms vary significantly. For each epoch we calculated the median absolute deviation (MAD) of the flux density ratios and identified six pointings (namely L233804, L346136, L346154, L346454, L347494 and L782679) with particularly poor integrated flux density values compared to the deep image (MAD values more than 1.5 times the median MAD). The actual image RMS values at shown in Fig. \ref{Fig:epoch_flux_densities}. The figure also shows these RMS values adjusted to reflect variations in bandwidth, integration time and flagging fraction (blue crosses), as well as the inverse of the source sky density within 30\% of the power primary beam (red circles). Here we see that the six epochs with poor integrated flux density values have consistently high RMS values and a low number of sources. 

\begin{figure}[htbp]
 \includegraphics[width=\linewidth]{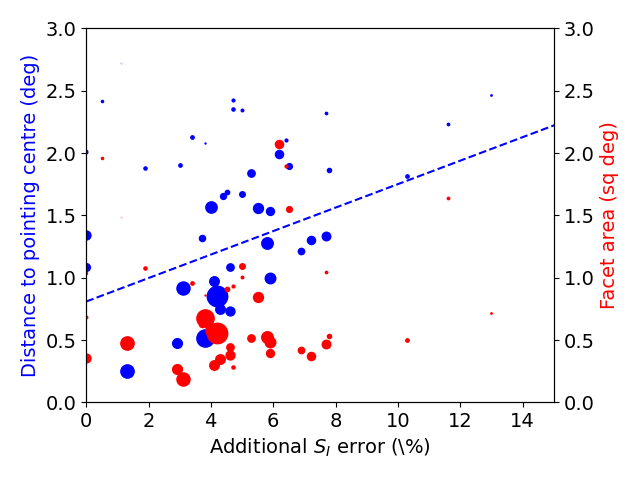}
 \vspace{-0.7cm}
   \caption{Additional fractional uncertainty in $S_I$ for each facet plotted against distance to the pointing centre (blue) and the facet area (red). The additional uncertainty is derived by comparing the distribution of ratios of integrated flux density measurements in each epoch to the deep integration for the 1447 sources detected in the 58 good epochs with a simple simulation using the same sources and errors but allowing for flux density scale offsets between pointings and facets. The consistency of our epoch image integrated flux density measurements decreases with increasing distance from the facet centre and the dashed blue line shows a linear fit to the data using the number of sources in the facet (proportional to marker size) for weighting.}
   \label{Fig:epoch_fluxquality_distance}
\end{figure}

If we exclude the six epochs identified as poor then 1447 sources are detected in every one of the remaining 58 images. These sources are located across the image with a mean of 33 sources per facet and a standard deviation of 22 among the facets. Using these sources we attempted to quantify flux density scale variations in the individual epoch images as a function of position. To do this we examined the distribution of the ratios of the integrated flux densities in the deep epoch to those in the individual epoch images for every facet. We compared these distributions to those derived from simple simulations using the same sources and errors but adding an additional fractional error to the source flux that is fixed for a given observation and facet (i.e. mimicking a flux density scale misalignment between the observations that is allowed to vary between facets). We then varied the amount of additional fractional error in the integrated flux density until the standard deviation of a Gaussian function fitted to the simulated distribution matches the real distribution. The results of this are shown in Fig. \ref{Fig:epoch_fluxquality_distance} which indicates that for regions further from the pointing centre there are generally larger discrepancies on the integrated flux density measurements between observations: a crude linear fit indicates an additional error on the integrated flux density of $\sigma_{S_I,A} = \frac{D-0.76}{10} S_I$ for $D>0.76$ where $D$ is distance from the pointing centre in degrees. This effect may be partly due to calibration issues (e.g. we note that facets further from the pointing centre also tend to be larger, implying that the ionospheric calibration may be less precise) or possibly uncertainty in the LOFAR beam model could play a role. The mean additional flux density scale uncertainty of all the facets is 5\%. This level represents the residual flux density scale uncertainty between individual epoch images after they have been aligned with the flux density scale of our deep image.

\section{Image analysis}
\label{sec:image_analysis}

Here we discuss several properties of the dataset to help demonstrate its scientific potential, its limitations and the improvements it provides upon previous work. We  evaluate the sensitivity and its limitations from confusion noise, and the ability of the data to discover variable and polarised sources and to probe faint diffuse emission.

\subsection{Comparison with \cite{Sabater_2021}}

\begin{figure}[htbp]
   \centering
   \includegraphics[width=\linewidth]{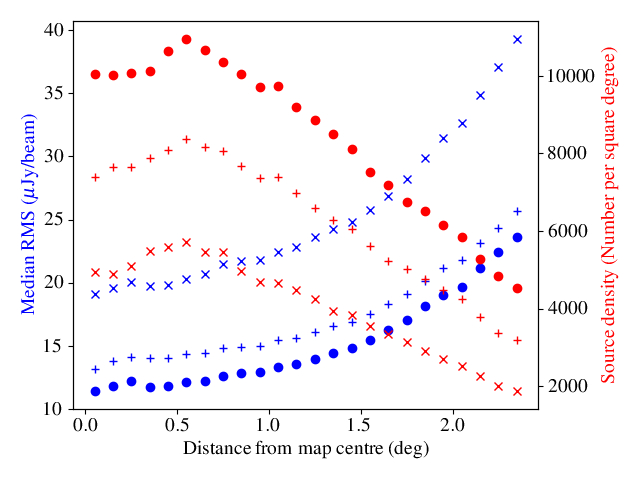}
   \vspace{-0.7cm}
   \caption{The source density (red) and the image RMS (blue) as a function of distance from the pointing centre. The x symbols show the values derived from the previous release of the LoTSS Deep ELAIS-N1 field (\citealt{Sabater_2021}), the $+$ symbols show the values derived from our image using the exact same \textsc{PyBDSF} parameters as \cite{Sabater_2021} and the circular symbols show the values derived using our revised \textsc{PyBDSF} strategy described in Sect. \ref{Sec:Sourcefinding}.}
   \label{Fig:Sabater2021comparison}
\end{figure}

For a simple quantitative comparison with \cite{Sabater_2021} we create a catalogue from our image using exactly the same \textsc{PyBDSF} parameters as used in this previous study. We examine the source density and image noise as a function of distance from the pointing centre and the results are shown in Fig. \ref{Fig:Sabater2021comparison}. The RMS in our new image is a mean factor of 0.68 times the RMS in the previous data release image with little radial dependence (minimum factor is 0.65 and maximum factor is 0.71) and generally the source density is 1.57 times higher (minimum of 1.43 and maximum of 1.71). Using the refined \textsc{PyBDSF} parameters which detect sources at a lower level of significance and better characterise extended structures we again significantly increase (by a mean factor of 1.34) the number of sources in the catalogue derived from our new image compared with using the old \textsc{PyBDSF} parameters on the same image. Furthermore, adopting this new approach means  additional faint sources are removed from the residual image from which the RMS is calculated. This results in a moderate decrease (mean factor of 0.87) in the measured RMS of our new image compared to when using the old  \textsc{PyBDSF} parameters on the same image. 

\subsection{Confusion noise}
\label{sec:confusion}

Confusion noise is a consequence of the population of faint sources with flux densities below the detection/deconvolution threshold. Assuming that this population consists solely of point sources then each undetected source will contribute to the image as the data PSF multiplied by the flux density of the source with both the source and associated sidelobes adding noise to the image. When a large number of sources contribute to the confusion noise then these contributions form a substantial background of noise that has a highly skewed and non-Gaussian distribution with a non zero mean (e.g. \citealt{Condon_2012}, \citealt{Zwart_2015}).

\begin{figure}[htbp]
 \includegraphics[width=\linewidth]{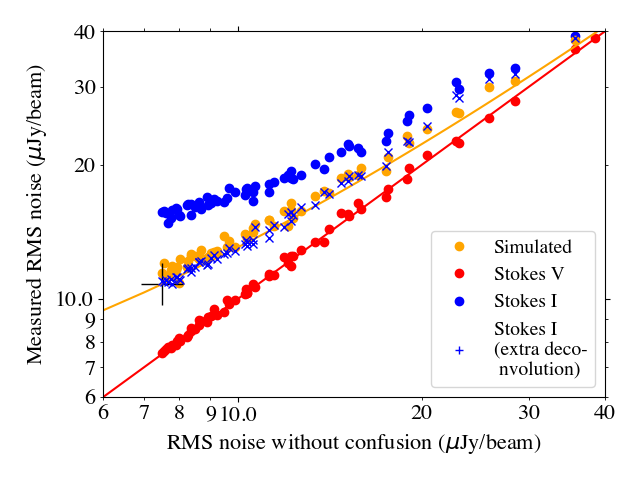}
 \vspace{-0.7cm}
   \caption{The impact of confusion noise at 144\,MHz and 6$\arcsec$ resolution. The blue crosses and circles show the noise measured from our stacked Stokes I residual images with and without image-plane deconvolution of sources that would be detected and deconvolved if the $uv$-data were imaged together. The orange circles show the measurements from our simulations which are fit with $\sigma_{I_{RMS}} = 0.9 \sigma_{V_{RMS}} + 5.7\times10^{-4} \sigma_{V_{RMS}}^2 + 3.96$. Here $\sigma_{I_{RMS}}$ is the measured Stokes I sensitivity (root mean square noise) and $\sigma_{V_{RMS}}$ is the sensitivity without confusion, both are in $\mu$Jy beam$^{-1}$.  In red are the measurements from our stacked Stokes V maps. The $y$-axis shows the total noise measured from stacked DDFacet residual images whereas the $x$-axis shows the thermal noise estimated from the weights of the Stokes V images for datasets included in that stacked image. The red line shows where these are equal and the large black cross is the sensitivity of our final ELAIS-N1 6$\arcsec$ resolution image.}
   \label{Fig:confusion_noise}
\end{figure}

As was already apparent in the work of \cite{Sabater_2021} the sensitivity of the $6\arcsec$ resolution LoTSS Deep images is impacted by confusion noise even at a sensitivity of 20\,$\mu$Jy beam$^{-1}$. To quantify the impact of confusion noise as a function of image depth we examine how the image noise decreases as a function of observing time and assess how it deviates from expectations, or comparable measurements, without confusion noise. To explore this it is prohibitively computationally expensive to reimage the large dataset with various amounts of integration time removed. Instead we stacked the individual epoch DDFacet residual Stokes I images to mimic different total integration times. To do this we randomly chose a number (2-64) of individual epoch images and stacked them together in the image plane using the weighted ($\sigma_{V_{RMS}}^{-2}$, where $\sigma_{V_{RMS}}$ is root mean square Stokes V image noise) mean which accounts for variation in the image quality from factors such as observing elevation and ionospheric conditions/calibration quality. The total noise on each stacked image was then estimated by finding the 68.27\% percentile of a function fitted to the distribution of the stacked DDFacet residual map noise pixels after removing regions from the image impacted by dynamic range (see Sect. \ref{sec:dynrange}). The thermal noise contribution to the total noise on each stacked image was found from Stokes V images stacked using the same epochs and weighting. In Fig. \ref{Fig:confusion_noise} we demonstrate that the noise properties measured from the stacked individual epoch Stokes V dirty maps (the data were processed before DDFacet had Stokes V deconvolution functionality) is approximately equal to the noise estimated from the weights of the individual images (the standard deviation of the ratio of the measured noise to the estimated noise is 0.02). This behavior is expected as the noise in these maps does not suffer from confusion, since sources are generally not circularly polarised and our levels of leakage are low (see Sect. \ref{Sec:StokesV}). 

The analysis is complicated by the fact that each individual epoch image is only deconvolved to a certain threshold (five times the local noise of typically $\sim70\mu$Jy beam$^{-1}$) and therefore sources below this threshold are present in our stacked DDFacet residual images and contribute to the confusion noise in those maps. This is different from how the deep image was created as there we used all the data during imaging and deconvolved to a far lower threshold (five times the local noise of approximately $\sim10.7\mu$Jy beam$^{-1}$). To account for this during the stacking of our DDFacet residual images we searched for sources that were present in the deep image unresolved source catalogue and if these sources were detected in the stacked DDFacet residual image with a significance exceeding 3 times the local noise (obtained from a weighted stacking of the individual epoch \textsc{PyBDSF} noise maps) then we subtracted them in the image plane, assuming that the sources are point-like and have the response of the PSF of the central facet (DDFacet calculates a PSF for each facet). We did this by subtracting the peak brightness multiplied by the normalised PSF centred on the source position. This subtraction means that these sources and associated sidelobes no longer contribute to the measured confusion noise in the stacked DDFacet residual image measurements thus approximately mimicking the joint deconvolution of the stacked data.

Our measurements showing the noise estimates on our stacked DDFacet residual images compared to the thermal noise estimates from the Stokes V images are shown in Fig. \ref{Fig:confusion_noise}. We also show the measurements from the stacked DDFacet residual images prior to the image-plane deconvolution to demonstrate the impact of that procedure.  In this plot we clearly see that as more DDFacet residual images are stacked the sensitivity continues to improve but that the measured stacked DDFacet residual image noise deviates further from the corresponding thermal noise estimate. This is because, even though the confusion noise is decreasing as more data are stacked due to the image-plane deconvolution removing detected sources from the stacked DDFacet residual and the number of undetected sources below the detection threshold decreasing, there is still a population of faint sources that remain in the stacked DDFacet residual images which form the confusion noise. 

\begin{figure}[htbp]
   \centering
   \includegraphics[width=\linewidth]{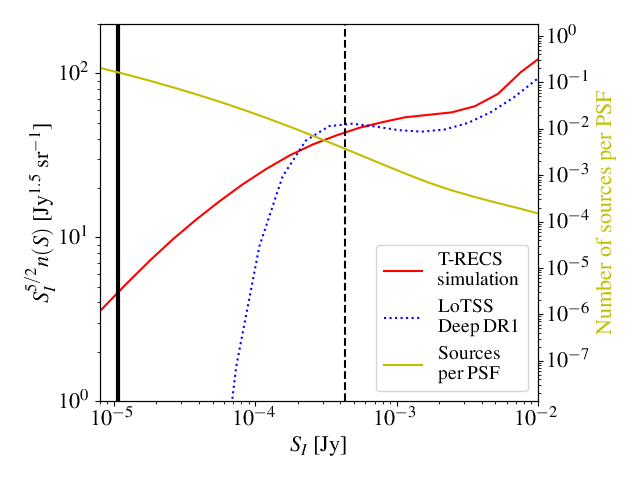}
   \vspace{-0.7cm}
      \caption{Euclidean normalised source counts at 144\,MHz ($y$-axis on the left) from previous surveys and simulations. The dashed blue line shows the counts from \cite{Mandal_2021} who used the previous LoTSS Deep fields data release. The solid red line shows the counts from a T-RECS  (\citealt{Bonaldi_2023}) simulation at 144\,MHz. The \cite{Mandal_2021} counts drop rapidly at low flux densities due to the limited depth of the previous LoTSS Deep data release images. Hence in our confusion noise simulations (Sec \ref{sec:confusion}) we have extrapolated the \cite{Mandal_2021} counts below 0.43\,mJy  (shown by vertical dashed black line) with the T-RECS counts. The yellow line shows the number of sources above a given flux density per 6$\arcsec$ restoring beam ($y$-axis on the right) and the thick vertical black line shows the sensitivity of our final deep ELAIS-N1 6\arcsec image. Source counts derived from LoTSS Deep DR2 will be presented in a future study.}
   \label{Fig:TRECS_source_counts}
\end{figure}

The shape of the curve in the Stokes I DDFacet residual image measurements in Fig. \ref{Fig:confusion_noise} can be compared to  simulations that attempt to reconstruct the confusion noise based on an assumed source population and the true image PSF.  At low flux densities there is uncertainty about the source population as it has not yet been measured but it can be estimated from simulations and existing counts. In Fig. \ref{Fig:TRECS_source_counts} we show a fit of the differential source counts normalised to a non-evolving Euclidean model derived from the previous LoTSS Deep release by \cite{Mandal_2021}. At low integrated flux densities ($\lesssim0.43$\,mJy) the normalised counts begin to rapidly decline but, given statistical limitations due to the depth of the previous data release, the uncertainty becomes increasingly large. Instead of following this rapid decline we extrapolate the source counts from \cite{Mandal_2021} at 0.43\,mJy assuming the number counts down to 1\,$\mu$Jy from the T-RECS simulated source population at 144\,MHz (\citealt{Bonaldi_2023}).

In our simulations we take the PSF of the central facet together with our approximation of the real source counts and for each source inject the appropriately scaled PSF into the stacked Stokes V map (i.e. the thermal noise map) at a random location. To mimic DDFacet residual images, we then remove sources (PSFs) that exceed a detection threshold (5$\sigma$) as these would have been deconvolved during the imaging and would no longer be in the DDFacet residual image.  We continue subtracting sources until the image noise converges.

In Fig. \ref{Fig:confusion_noise} we plot the results of our simulations together with the measurements from the stacked DDFacet residual images as well as the derived noise estimate on our final jointly deconvolved map where we have used the weights from the Stokes V image stacked map to estimate the thermal noise. Through both our image-plane deconvolved stacked images and our simulations we are able to reproduce the noise achieved in the final jointly deconvolved map (implying the source counts provide comparable confusion noise to what is observed) as well as the trend for increasing deviations from thermal noise with decreasing sensitivity. Our final image reaches 10.7\,$\mu$Jy beam$^{-1}$ sensitivity where approximately half (defined as $\sigma_{I_{CONF}} = \sqrt{\sigma_{I_{RMS}}^2-\sigma_{V_{RMS}}^2}$) of this is from confusion noise, $\sigma_{I_{CONF}}$ (i.e. we would have reached an RMS noise level of 7.5\,$\mu$Jy beam$^{-1}$ if there were no confusion).  Extrapolating our measurements we find that if we were to double the integration time, confusion would become even more dominant - an estimated total noise of 8.8\,$\mu$Jy beam$^{-1}$ or $\sigma_{V_{RMS}}$=5.3\,$\mu$Jy beam$^{-1}$ (i.e. noise without confusion)  and confusion noise of $\sigma_{I_{CONF}}$=7.0\,$\mu$Jy beam$^{-1}$.

\subsection{Compact source variability}
\label{Sec:variablity}

We first constructed a catalogue containing all isolated unresolved sources in the field that we wish to search for variability. To do this we begin by searching for sources that are only detected in any of the 58 individual epoch images (excluding the six poor quality epochs identified in Sect. \ref{Sec:indiv_quality}) and not in the deep image. For this we filter each individual epoch catalogue so that it just contains $\frac{S_I}{\sigma_{S_I}}>5$ unresolved isolated sources, where again `unresolved' is defined as derived in Sect. \ref{sec:source_extensions} together with having \textsc{PyBDSF} source code `S' and `isolated' is defined as having no nearest neighbours within 15$\arcsec$. We cross-match each of these filtered individual catalogues with the deep catalogue using a cross matching radius of 4$\arcsec$ and find that there are only two unresolved isolated sources detected in the individual fields (both in L811736) that are not also in the deep catalogue, but one of these is clearly an artefact and associated with a nearby (0.1$^\circ$) bright (2.8\,Jy) source so it is removed.

We then filtered the deep catalogue (154,952 entries plus 1 entry for the source found just in L811736) using the same criteria as above which removes 83,722 due to the SNR cut, a further 11,924 due to the nearest neighbour cut, 12,110 from the S\_Code cut, 4096 are removed by the size cut, and 3365 due to dynamic range cuts. This left 39,736 sources that we can search for variability. For each source and each epoch we recorded the catalogued $S_I$ and $S_P$ and the corresponding errors, which are all scaled by the individual field scaling factors (derived in Sect. \ref{Sec:indiv_quality}) to align with the deep field catalogue. If no source was detected in a particular individual epoch catalogue, we instead derived $S_I$ and $S_P$ and associated errors from the primary beam corrected individual epoch images. This was done by taking the pixel value corresponding to the source position in the deep catalogue as well as the local RMS around this position. Both values were scaled by the appropriate individual field scaling factor to align with the deep field catalogue. For all sources we also recorded the sum of the pixels within the \textsc{PyBDSF} deep image mask island for that particular source scaled by the individual field scaling factors ($S_{I,ap}$) and associated error ($\sigma_{ap,tot}$). Out of the 39,736 sources, 37,676 of these are detected in at least one of the 58 individual epoch observations whilst only 1447 are detected in all epochs. The 2060 sources not detected in any of the individual epochs were not examined further.

For each of the 37,676 sources in this cross matched catalogue we calculated the commonly used variability parameters $V$ and $\eta$ (see e.g. \citealt{Rowlinson_2019}) which are proxies for the amount of variability and the variability significance respectively. These 
 are defined as:
\begin{equation}\label{eq:resolved_sources}
V = \frac{1}{\overline{S}}\sqrt{\frac{N}{N-1}(\overline{S^2}-\overline{S}^2}),
\end{equation}
\begin{equation}
\eta = \frac{N}{N-1}\left({(\overline{\omega S^2}-\frac{\overline{\omega S}^2}{\bar{\omega}}}\right)
\end{equation}
where $V$ is the ratio of the standard deviation of the flux densities to the mean flux density, $\eta$ is the reduced $\chi^2$ relative to a model with constant flux density, $N$ is the number of measurements (58), $\omega=1/\sigma_{I,tot}^2$, overbars denote mean and an additional uncertainty that we define as 5\% of the flux density (see Sect. \ref{Sec:indiv_quality}) is added in quadrature to the $\sigma_{S_I}$ derived by \textsc{PyBDSF} to give the total error on integrated flux density ($\sigma_{S_I,tot}$).

If a source is one of the 1447 detected in all epochs we calculated the variability parameters using the catalogued epoch $S_I$, $S_P$ and $S_{I,ap}$ values as $S$. Additionally, we repeated the variability calculation 1000 times drawing $S$ from a Gaussian distribution centred on $S_I$ with width $\sigma_{S_I,tot}$ and a further 1000 calculations drawing from around each of $S_{I,ap}$ and $S_P$ instead of $S_I$. For each flux density type we then conservatively selected the lowest derived $V$ and $\eta$ for that source from all the calculations (i.e. least amount of variability given the measurements).

Most sources, however, are only detected in a few epochs. For these sources we performed the same analysis as above but for the epochs where the source is not detected we drew $S$ from a uniform probability distribution between 0 and the upper limit (e.g $5\times \sigma_{I,tot}$, $5\times \sigma_{P,tot}$ or $5\times \sigma_{ap,tot}$). We set the error on the flux density values of these undetected sources as being the standard deviation of a population of flux densities drawn from a uniform distribution between 0 and the upper limit. Again for these sources we conservatively selected the lowest derived $V$ and $\eta$ from all the calculations. In total we thus have three minimum $V$ and $\eta$ values for each source corresponding the $S_I$, $S_P$ and $S_{I,ap}$ measurements.

\begin{figure}[htbp]
 \includegraphics[width=\linewidth]{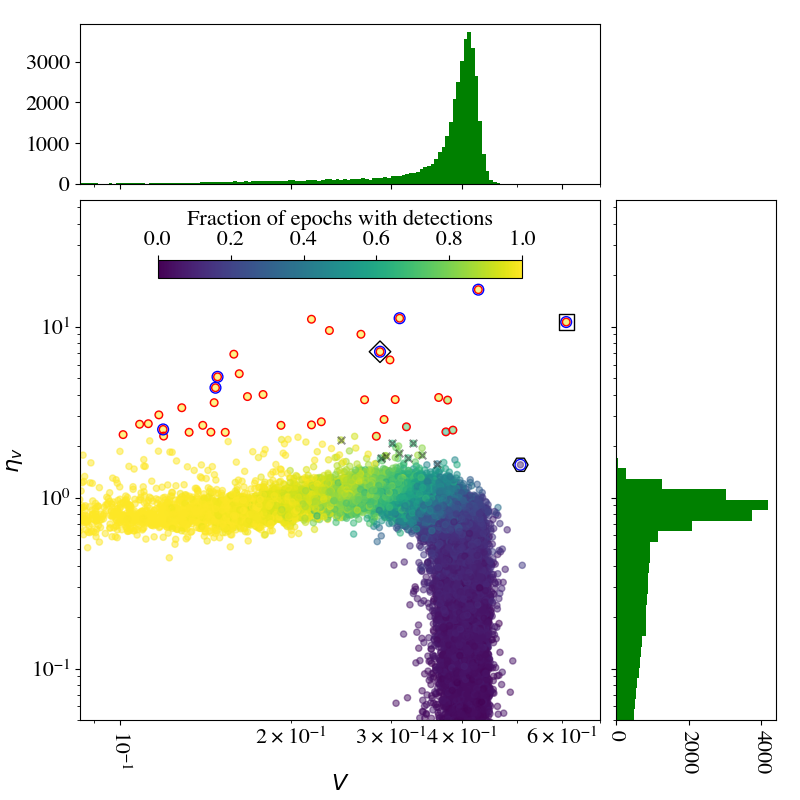}
 \vspace{-0.7cm}
   \caption{The derived variability parameters $V$ and $\eta$ for the 37,676 sources detected in at least one epoch. The marker colour shows the fraction of epochs in which the source is detected. More variable sources have high values of $V$ and $\eta$. The 38 candidate variable sources are circled in red and the sources shown in Fig. \ref{Fig:example_variable_sources} are circled in blue. The box, diamond and hexagon show Cr Draconis, PSR J1552+5437 and GJ 625 respectively. The grey x's show additional candidate variable sources identified when using Gaussian rather than uniform probability distributions for non detections (see text).}
   \label{Fig:all_variable_sources}
\end{figure}

\begin{figure}[htbp]
 \includegraphics[width=\linewidth]{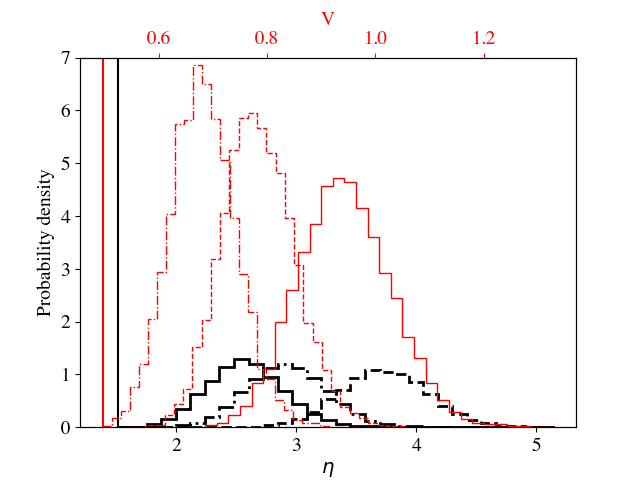}
 \vspace{-0.7cm}
   \caption{The probability density distributions of the derived variability parameters $V$ and $\eta$ for GJ 625. In black we show $\eta$ with the solid, dashed and dot dashed lines corresponding to the $S_I$, $S_P$ and $S_{I,ap}$ measurements respectively. The red shows the $V$ distributions using the same line style.  The red and black vertical lines show the lowest derived $V$ and $\eta$ parameters (i.e. least variation) that we adopt.}
   \label{Fig:variability_vales_GJ625}
\end{figure}

Fig. \ref{Fig:all_variable_sources} shows the lowest of the three minimum variability parameter sets for each of the 37,676 sources we have examined - i.e. these are the lowest values of both the variability parameters for each source that we are able to derive given the measurements (see Fig. \ref{Fig:variability_vales_GJ625} for an example). They are colour coded by the fraction of epochs a particular source is detected in. For demonstration purposes we have also included the well known flare star binary CR Draconis (see e.g. \citealt{Callingham_2021b}) even though it was excluded from the sample as it is in a dynamic-range limited area of the map. We also included the millisecond pulsar PSR J1552+5437 (\citealt{Pleunis_2017}) which was excluded due to \textsc{PyBDSF} fitting the emission with multiple Gaussians. As can be seen from the shape of the distribution the sources that are detected in almost all epochs show a slightly increasing variability significance ($\eta$) with variability ($V$). This is a consequence of the 5\% fractional error that we have placed on the flux density - fainter sources tend to have larger variability ($V$) because the source fluxes are dominated by measurement errors and the flux density changes significantly from epoch to epoch. For brighter sources that are detected in almost all epochs the fractional uncertainty dominates over the measurement errors. For these sources we find rapidly decreasing $V$ with increasing flux density whereas $\eta$ only slightly decreases with increasing flux density. If the calculations are repeated with a 10\% fractional error on the flux density this trend of increasing $V$ with $\eta$ for sources that are detected in most epochs becomes much stronger as $\eta$ more rapidly decreases with increasing flux density. The weak trend we see with a 5\% fractional uncertainty gives confidence that this fractional uncertainty approximately reflects the correct value for the data under the assumption that sources that are detected in more epochs are genuinely of similar variability to sources detected in fewer epochs.

Sources that are detected in few of the epochs have a large variability as for each undetected epoch we are randomly assigning the flux density between 0 and the 5$\sigma$ upper limit. However, the variability significance ($\eta$) is low because the assigned flux densities in undetected epochs have a high degree of uncertainty (the standard deviation of a population of flux densities in the range between 0 and the upper limit).

\begin{figure*}[htbp]
 \includegraphics[width=0.24\linewidth]{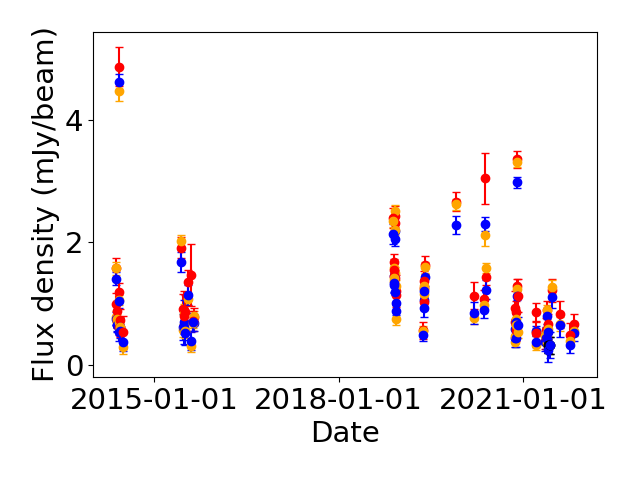}
  \includegraphics[width=0.24\linewidth]{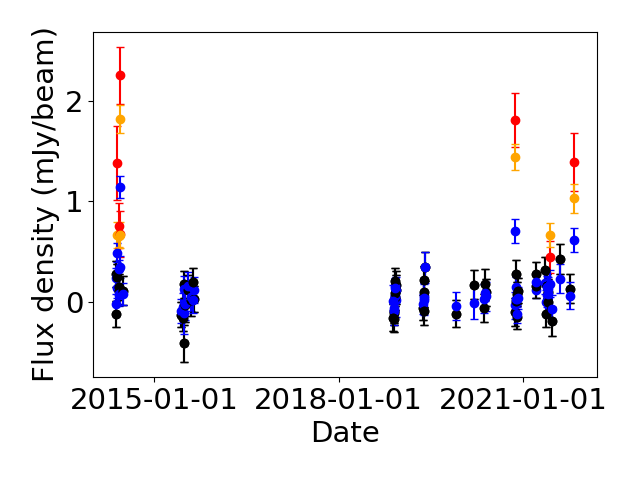}
\includegraphics[width=0.24\linewidth]{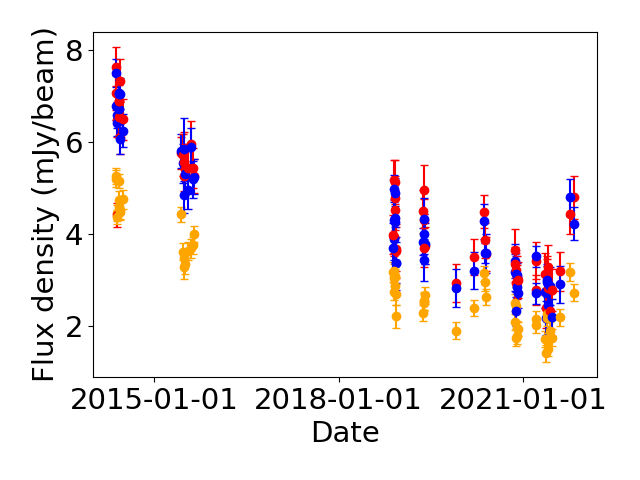}
   \includegraphics[width=0.24\linewidth]{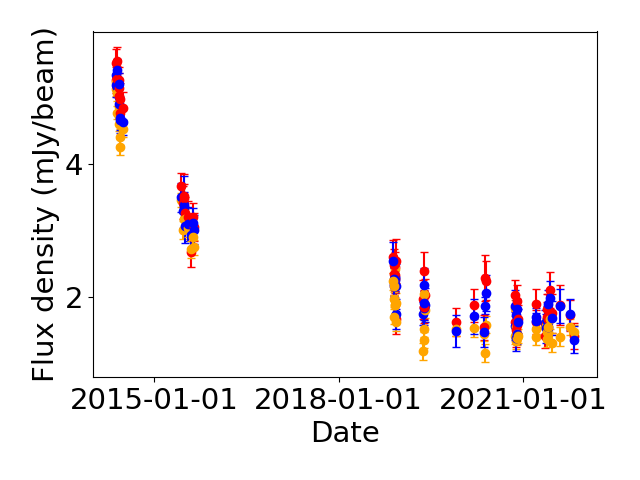} \\
   \includegraphics[width=0.24\linewidth]{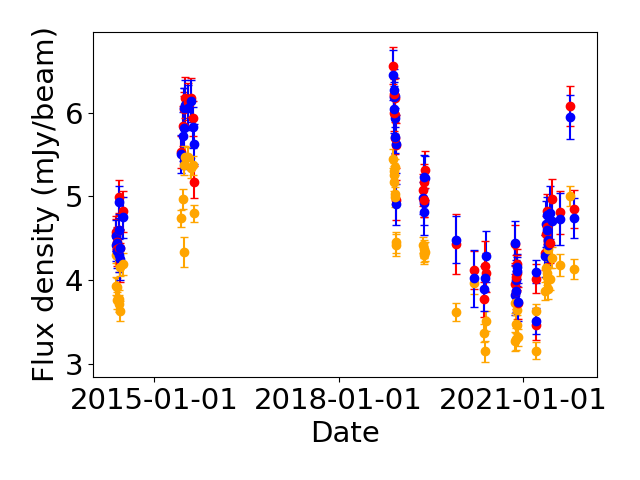}
   \includegraphics[width=0.24\linewidth]{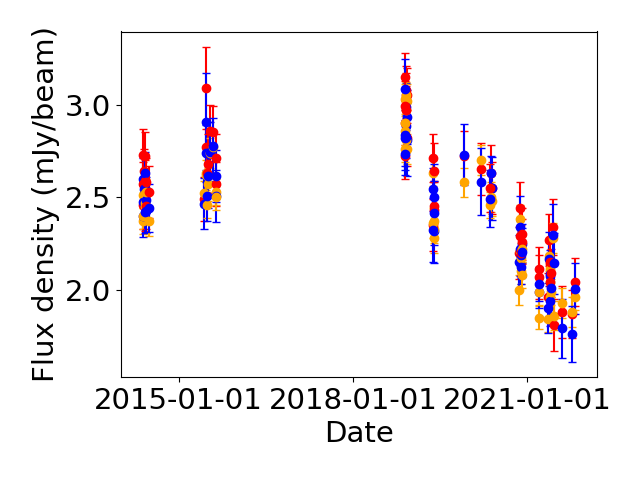}
   \includegraphics[width=0.24\linewidth]{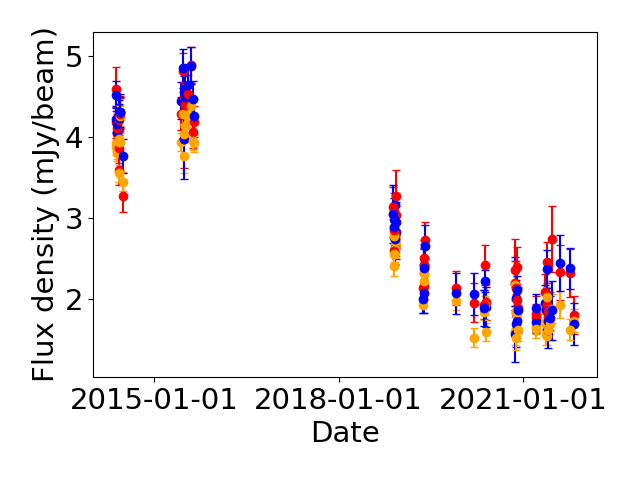}
   \includegraphics[width=0.24\linewidth]{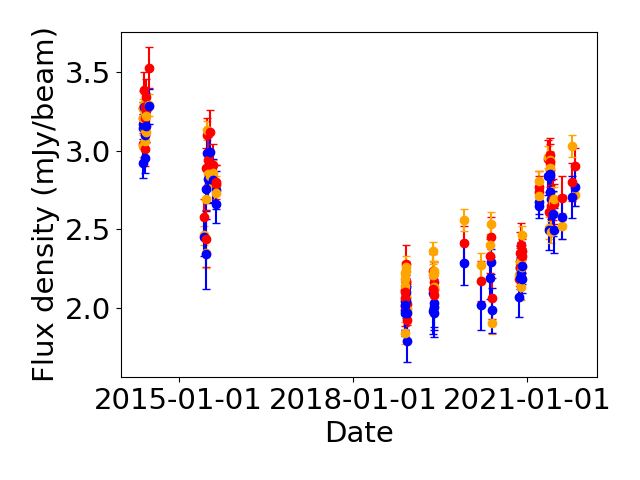}
   \vspace{-0.2cm}
   \caption{The $S_I$ (red), $S_P$ (orange), $S_{I,ap}$ (blue) and pixel values for non detections (black) as a function of time for a selection of variable sources highlighted in blue in Fig. \ref{Fig:all_variable_sources}. Clockwise from the top left the panels show CR Draconis, GJ 625, PSR J1552+5437 and the others are classified as a RQAGN, LERG, SFG, LERG and unclassified, respectively, by \cite{Best_2023}.}
   \label{Fig:example_variable_sources}
\end{figure*}

Outliers in Fig.\ref{Fig:all_variable_sources} with an unusually high $\eta$ are our candidate variable sources. These are distributed across the entire image (see Fig. \ref{Fig:ELAIS_image}) and have a range of different levels of variability ($V$). The 38 sources with $\eta$ above the 99.9 percentile are circled in red and those circled in blue are the six example sources that we show in Fig. \ref{Fig:example_variable_sources}. The sources with high $\eta$ include the flare star CR Draconis (see e.g. \citealt{Callingham_2021b} and the millisecond pulsar PSR J1552+5437 (see e.g. \citealt{Pleunis_2017}). The small red dwarf star with an exoplanetary companion GJ 625 is a prominent outlier in $V$ albeit with a lower $\eta$ (circled blue but not red). The other 36 sources with high $\eta$ were all classified as part of the \cite{Best_2023} LoTSS Deep fields DR1 classification effort and were classed as 2 high-excitation radio galaxies (HERGs), 17 low-excitation radio galaxies (LERGs), 3 radio-quiet AGN (RQAGN), 11 star-forming galaxies (SFGs) and 3 unclassified. Furthermore, we note that all 36 of these sources are highly compact as they  are detected in the 0.3$\arcsec$ resolution LOFAR image (\citealt{deJong_2024}) and have measured major axis below 1.2$\arcsec$ in that catalogue. If the emission is genuinely compact then a possibility is that refractive scintillation plays a significant role (e.g. \citealt{Bell_2019}) but the modulation index (here defined as $M= \frac{\sigma_{S,epochs}}{\overline{S}}$, where $\sigma_{S,epochs}$ is the standard deviation of the epoch flux density measurements and $\overline{S}$ is the mean flux density measurement from the epochs) we measure for these 36 sources using all our data is  comparably high (between 12\% and 55\%). Further investigation is required to more precisely characterise the variability and confirm its nature.

The cautious approach we have taken to search for variable sources in the field uses three different ways of measuring the flux density and searches for the minimum levels of variability given the measurements and errors. This approach was used to demonstrate that substantial populations of variable sources exist in these data and are identifiable despite the variations in the image characteristics between epochs (e.g. Fig. \ref{Fig:epoch_flux_densities}). There are many refinements that could be made to the procedure to identify larger numbers of variable sources in future studies, such as better characterisation of local variations in image quality, expanding the search to cover more sources, making fuller use of the image values, improving the identification of outliers in the variability plot and exploring different time periods (combining epochs or within epochs). For example, we can make more use of the data if we simply repeat the calculations but for epochs without detections we instead use Gaussian probability distributions centred on the image measurements of $S_I$, $S_P$ and $S_{I,ap}$ and with standard deviations set to the corresponding errors. By doing this we find that for sources with few epoch detections typically the $\eta$ increases slightly ($\sim$15\%) whilst the distribution in $V$ broadens substantially and stretches to far higher values ($\sim1.0$). Overall though the sources selected as most significantly variable are largely the same as those previously identified (29 out of 38) with only the candidate variable sources close to the 99.9 percentile boundary shifting one way or the other. The main difference was that this approach resulted in more low signal-to-noise candidate variables that are detected in just a few epochs (marked with grey x's in Fig. \ref{Fig:all_variable_sources}), whereas the uniform probability distribution approach to non detections led to more candidates from sources that are detected in the  majority of the epochs. 

To allow for examination of variability within these data we include in this data release a catalogue containing the flux density and error measurements from each epoch in addition to the images from individual epochs and the derived minimum $\eta$ and $V$ values.

\subsection{Stokes V emission}
\label{Sec:StokesV}

There are three known circularly polarised sources within the field, CR Draconis, GJ 625 and PSR J1552+5437. The circularly polarised emission from these sources was identified either in this ELAIS-N1 LOFAR dataset by \cite{Callingham_2021a}, \cite{Callingham_2021b} and \cite{Sabater_2021} or in LoTSS by \cite{Callingham_2023}. Here we search for additional Stokes V emission at the locations of each of the 39,676 unresolved sources identified in Sect. \ref{Sec:variablity} and we also measure the emission from the three known Stokes V emitters to demonstrate our methodology. We used several search techniques to try to take advantage of the multi-epoch dataset whilst recognising that the handedness of Stokes V sources can vary with time and changes would cause the signal to flip from positive to negative or vice versa in the Stokes V images.

\begin{figure}[htbp]
 \includegraphics[width=\linewidth]{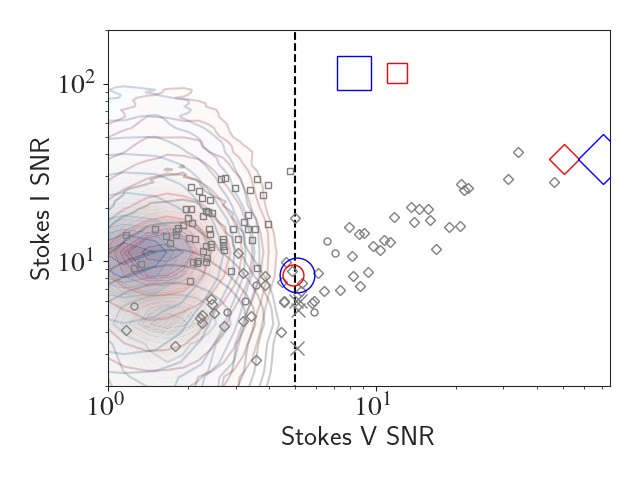}
\vspace{-0.7cm}
   \caption{The derived fractional circular polarisation for the  37,676 unresolved, isolated sources detected in at least one epoch in a region of the map not impacted by dynamic range limitations. CR Draconis and PSR J1552+5437 are also included for demonstration purposes even though they fall outside the selection criteria. The grey (570,944 measurements), red and blue contours show the probability density function for the individual epoch measurements, stacked measurements and stacked absolute value measurements respectively. The lowest contours contain 99\% of the measurements. The diamonds, circles, box and x's show CR Draconis, GJ 625, PSR J1552+5437 and other detections above 5 SNR respectively. The smaller grey symbols represent measurements from the individual epochs, the medium sized red symbols are from the stacked images and the large blue symbols are from the stacked absolute value images. Details of the SNR derivations are given in Sect. \ref{Sec:StokesV}.}
   \label{Fig:all_stokesV_sources}
\end{figure}

Firstly, we used \textsc{PyBDSF} to create noise images for each of our Stokes V individual epoch images and used these to measure the Stokes V noise at the positions of each source in each epoch. We then searched for Stokes V emission by finding the maximum absolute value of the Stokes V image in the pixels that lay both within the appropriate deep image Stokes I \textsc{PyBDSF} mask island and within 4$\arcsec$ of the catalogued deep image position. This search was performed for every epoch image in which the source was detected in Stokes I giving a total of 570,944 Stokes V measurements. For consistency we scaled our Stokes V measurements and errors by the individual field scaling factors that were derived to align the individual epochs with the deep catalogue. In Fig. \ref{Fig:all_stokesV_sources} we show our measured Stokes V signal divided by noise and the corresponding Stokes I epoch measurements (\textsc{PyBDSF} catalogued $S_P$ and $\sigma_{S_P}$). In total 6 different sources are detected in Stokes V with a SNR exceeding 5. The two most prominent detections are of the flare star CR Draconis (33 detections with Stokes V  $5.3 < SNR < 46.2$) and the small red dwarf star GJ 625 (3 detections with Stokes V $5.9 < SNR < 7.0$). The pulsar PSR J1552+5437 is detected at a maximum significance of 4.8 whereas another 4 sources have Stokes V SNR levels very slightly exceeding 5 (all less than 5.2) in a single epoch, these are sources with identities ILTJ161318.70+535057.6 (L798164), ILTJ161227.28+544553.2 (L686962), ILTJ160926.33+564927.2 (L230461) and ILTJ160859.34+531359.5 (L816344). These 4 sources have no obvious highly circularly polarised counterparts such as stars or known pulsars and may be false positives. To assess this we approximated the expected number of false positives by finding the number of noise pixels exceeding 5 SNR in the Stokes V cutout images outside of the deep field \textsc{PyBDSF} mask regions (i.e. away from all sources). Between all the epoch images we found 158  $>5$ SNR Stokes V pixels out of 0.4 billion pixels searched away from sources (i.e. $6.6\times10^{-5}\%$) and all had SNR $<6$. By comparison, if we exclude CR Dra, GJ 625 and PSR J1552+5437, then in our search for Stokes V emission in pixels associated with detected sources we identified 12 SNR $>$5 Stokes V pixels out of 21 million pixels searched (i.e. $5.6\times10^{-5}\%$). Hence our detection rate at source locations is approximately equal to our false positive rate and we conclude that our 4 additional detections of Stokes V sources are likely spurious.

In order to see if other detections could be made we searched for signals in stacked Stokes V cutout images of small areas around each individual source. We did this via two approaches. Firstly, we simply stacked the images using a weighted average (using the Stokes V \textsc{PyBDSF} noise image and inverse-variance weighting) and characterising the significance of detections in the same way as was done for the individual epoch sources. This approach works well for sources that do not change handedness but signals from sources that change handedness would be diluted. In this approach we detected CR DRa and PSR J1552+5437 at high significance with SNR of 50.7 and 12.0 respectively. No other sources were detected above a SNR of 5.0 with the next most significant being GJ 625 with an SNR of 4.98. In the second approach the stacking was done using an inverse-variance weighting of the absolute values of the Stokes V cutout images from all the epochs for each source, where the absolute value was used to account for possible sign changes between epochs. To  quantify the significance of the measurements from the Stokes V stacked cutout images we evaluated the distribution of pixels of the full field stacked absolute value Stokes V epoch images. The pixels are well described by a beta distribution and we can quantify significance using confidence intervals, e.g. the confidence intervals of 68.27\%, 95.45\% and 99.73\% corresponding to an SNR of 1, 2 and 3 respectively. The results from this analysis are also shown in Fig. \ref{Fig:all_stokesV_sources} with the Stokes V stacked SNR compared to the deep field $\frac{S_I}{\sigma_{S_P}}$. The stars CR Dra and GJ 625 were detected with moderate to high significance as was the pulsar PSR J1552+5437 (72.0, 5.5, 8.4 respectively) but no other sources were detected above a significance of 5.0. Repeating the analysis but instead only considering epochs in which sources are detected in Stokes I, we were able to detect the known Stokes V emitters at improved SNR (77.5, 13.2, 10.5 for CR Dra, GJ 625 and PSR J1552+5437 respectively) but again we found no other sources with a significance exceeding 5.0.

Finally we attempted to search for a statistical detection from the 39,674 sources without obvious Stokes V emission (i.e. excluding CR Dra, GJ 625 and PSR J1552+5437 which are genuine Stokes V emitters) through another stacking exercise. We use the Stokes V cutouts for all epochs for all of the 39,674 sources. We then calculated the inverse variance weighted average of the absolute value of all the Stokes V cutout images. To assess the significance of any signal in the stacked on-source Stokes V image we use offset cutouts taken around each source and add these together in the same way. We fitted a beta distribution to the stacked offset cutout image and again quantified significance with confidence intervals. In addition to this we also created an artificial leakage image where we took the Stokes I flux density value in a particular epoch convolved to the 6$\arcsec$ restoring beam and multiplied that by leakage levels in the range of 0.01\% to 1\% (previously for bright sources in LoTSS-DR2 the leakage level was found to be 0.056\%; \citealt{Shimwell_2022}), added it to the stacked offset Stokes V image and performed the same weighted stacking. Through this analysis we found no Stokes V signal in our stacked on-source image (SNR of 0.5) but we are able to detect the artificial leakage signals at an SNR exceeding 5.0 when the leakage is greater than 0.045\%. 

To summarise, three sources (CR DRa, GJ 625 and PSR J1552+5437) are detected clearly in Stokes V using several different approaches but no other convincing detections are made through either searching individual observations or stacking. A search for a statistical detection of Stokes V emission from all unresolved sources in the field also yielded a null result, implying that the typical level of emission from these sources when combined with leakage is less than 0.045\%. 

\section{Public data release}
\label{sec:data_release}

The images and associated catalogues presented and analysed in this publication are publicly available and can be accessed via the LOFAR surveys webpage\footnote{https://lofar-surveys.org/}. This consists of the following:

\begin{itemize}
\item A full depth (10.7\,$\mu$Jy beam$^{-1}$) high (6$\arcsec$) resolution Stokes I image out to the 30\% power point of the LOFAR primary beam. 
\item A catalogue derived from the full depth 6$\arcsec$ resolution image that contains 154,952 radio sources and the corresponding 182,184 Gaussian components from \textsc{PyBDSF} (version 1.10.3). The columns are those provide by default by \textsc{PyBDSF} and include position, integrated flux density, peak brightness, size and estimated statistical errors on all parameters (i.e. not including radial effects or additional uncertainties on astrometry or flux density scale which should be added as appropriate).
\item High (6$\arcsec$) resolution Stokes I and Stokes V individual epoch images.
\item A catalogue derived from the 6$\arcsec$ individual epoch images that contains 36,676 unresolved isolated sources with flux density measurements from each epoch and derived variability parameters.
\end{itemize}

\section{Summary}
\label{sec:summary}

We have described the second and final 6$\arcsec$ resolution data release for the LoTSS Deep ELAIS-N1 field at 144\,MHz. Our image contains 154,952 sources (consisting of 182,184 Gaussian components) in a 24.53 square degree region and reaches depths of 10.7\,$\mu$Jy beam$^{-1}$. The image is amongst the most sensitive radio wavelength images ever released over such a large area when accounting for the typical spectral index of radio sources. For comparison, assuming a typical spectral index of -0.8 (\citealt{Bohme_2023}) our sensitivity is equivalent to 1.85\,$\mu$Jy beam$^{-1}$ at 1284\,MHz which is comparable to the depth of the MIGHTEE survey which will span over 20 square degrees at that frequency (\citealt{Heywood_2022} and \citealt{Hale_2024}). At higher frequencies deeper images have been produced for small areas. For example, the COSMOS-XS survey covered 180 square arcminutes and reached a depth of 0.53\,$\mu$Jy beam$^{-1}$ at 3\,GHz (\citealt{Vlugt_2021}) which is a factor of 1.8 deeper than the extrapolated LoTSS Deep sensitivity at this frequency (0.97\,$\mu$Jy beam$^{-1}$), albeit over 2 orders of magnitude less sky area.

In this paper we have characterised general properties of the LoTSS Deep ELAIS-N1 image finding that 9\% of the sources appear to be resolved and 7.4\% of the area is affected by dynamic range limitations. The flux density scale is thought to be good to better than 9\% and the standard deviations of RA and Dec offsets between our catalogue and Pan-STARRS DR2 are less than 0.2$\arcsec$ for high significance detections. We examined the contribution of confusion noise, finding that approximately half of the image noise is due to confusion and demonstrating that resolution is the most critical avenue to decreasing the noise level further. We searched for variable sources in the field and identified the 39 sources that are variable with high significance.  We also searched for Stokes V emitting sources and did not detect any previously unknown sources but found three detections of  known circularly polarised sources. We limited the circular polarisation signal of the population of our defined unresolved sources to be less than 0.045\% of the total intensity signal including leakage.

The data from this release are publicly available and can be accessed via the collaboration’s webpage. This includes deep images, individual epoch images and source catalogues. Observations are now complete for three other LoTSS Deep fields and publications examining the 6$\arcsec$, 0.3$\arcsec$ and polarisation properties are forthcoming. These will make use of up to $\sim$400\,hrs data for each field which is approximately a factor of four larger than was utilised for LoTSS Deep DR1. Future work will also present multi-wavelength identifications and characterisation for the radio sources detected in the LoTSS Deep fields and through the William Herschel Telescope Enhanced Area Velocity Explorer survey of LOFAR selected sources (WEAVE-LOFAR; \citealt{Smith_2016}) we will obtain optical spectroscopy for almost all sources in the central regions of these fields.

\section{Data availability}

All data products described in Sect. \ref{sec:data_release} are available on the LOFAR surveys webpage: \href{https://lofar-surveys.org/}{https://lofar-surveys.org/}. A table (Tab. S1) providing details of the individual observations and associated images is available via Zenodo, at \href{https://doi.org/10.5281/zenodo.14603969}{https://doi.org/10.5281/zenodo.14603969}.

\begin{acknowledgements}
This paper is based (in part) on data obtained with the LOFAR telescope (LOFAR-ERIC) under project code LC2\_024, LC4\_008, LT10\_012 and LT14\_003. LOFAR (\citealt{vanHaarlem_2013}) is the Low Frequency Array designed and constructed by ASTRON. It has observing, data processing, and data storage facilities in several countries, that are owned by various parties (each with their own funding sources), and that are collectively operated by the LOFAR European Research Infrastructure Consortium (LOFAR-ERIC) under a joint scientific policy. The LOFAR-ERIC resources have benefited from the following recent major funding sources: CNRS-INSU, Observatoire de Paris and Université d'Orléans, France; BMBF, MIWF-NRW, MPG, Germany; Science Foundation Ireland (SFI), Department of Business, Enterprise and Innovation (DBEI), Ireland; NWO, The Netherlands; The Science and Technology Facilities Council, UK; Ministry of Science and Higher Education, Poland.

This research made use of the Dutch national e-infrastructure with support of the SURF Cooperative (e-infra 180169) and NWO (grants 2019.056 \& 2023.036). The Jülich LOFAR Long Term Archive and the German LOFAR network are both coordinated and operated by the Jülich Supercomputing Centre (JSC), and computing resources on the supercomputer JUWELS at JSC were provided by the Gauss Centre for Supercomputing e.V. (grant CHTB00) through the John von Neumann Institute for Computing (NIC). 

This research made use of the University of Hertfordshire high-performance computing facility and the LOFAR-UK computing facility located at the University of Hertfordshire and supported by STFC [ST/P000096/1], and of the Italian LOFAR IT computing infrastructure supported and operated by INAF, and by the Physics Department of Turin university (under an agreement with Consorzio Interuniversitario per la Fisica Spaziale) at the C3S Supercomputing Centre, Italy.

CLH acknowledges support from the Leverhulme Trust through an Early Career Research Fellowship. CLH also acknowledges support from the Oxford Hintze Centre for Astrophysical Surveys which is funded through generous support from the Hintze Family Charitable Foundation. PNB and RK are grateful for support from the UK STFC via grants ST/V000594/1 and ST/Y000951/1. AB acknowledges financial support from the European Union - Next Generation EU. MJH and DJBS acknowledge support from the UK Science and Technology Facilities Council (STFC) via grants ST/V000624/1 and ST/Y001249/1. RJvW acknowledges support from the ERC Starting Grant ClusterWeb 804208. MB acknowledges funding by the DFG under Germany's Excellence Strategy – EXC 2121 ``Quantum Universe'' – 390833306. FdG acknowledges support from the ERC Consolidator Grant ULU 101086378. KJD acknowledges support from the STFC through an Ernest Rutherford Fellowship (grant number ST/W003120/1). SI acknowledges the ASPIRE programme at the Anton Pannekoek Institute for Astronomy. IdR and AR acknowledge support from the NWO Aspasia grant (number 015.016.033). IdR, AR, and JMGHJdJ acknowledge support from the NWA CORTEX grant (NWA.1160.18.316) of the research programme NWA-ORC which is (partly) financed by the Dutch Research Council (NWO). D.G.N. acknowledges funding from Conicyt through Fondecyt Postdoctorado (project code 3220195). LKM is grateful for support from a UKRI FLF [MR/T042842/1]. IP acknowledges support from INAF under the Large Grant 2022 funding scheme (project “MeerKAT and LOFAR Team up: a Unique Radio Window on Galaxy/AGN co-Evolution”).  FS appreciates the support of STFC [ST/Y004159/1]. 

\end{acknowledgements}

\label{lastpage}

\newpage

 \begin{appendix}
\section{\textsc{PyBDSF} Source detection}
\label{appendix_pybdsf}
\setlength\parindent{14pt}

Below we detail the multi-step \textsc{PyBDSF} process that we used to generate the catalogue for the ELAIS-N1 deep image.
\begin{itemize}
    \item Firstly, we run \textsc{PyBDSF} over the image using a similar set up to previous LoTSS Deep data releases that makes use of wavelet decomposition to better fit emission of different scales. However, as we want to remove sources from the image to be able to more accurately measure RMS, we use a 4$\sigma_{thresh}$ peak signal-to-noise ratio (SNR) source threshold (using the \textsc{PyBDSF} parameter \texttt{thresh\_pix}). From this we save the residual image generated (we denote this as \texttt{resid\_map\_1}). This uses the \textsc{PyBDSF} command:
\end{itemize}

\begin{verbatim}
bdsf.process_image(image, detection_image=
    app_image, thresh_isl=3.0, thresh_pix=4.0, 
    rms_box=(150,15), rms_map=True, mean_map=
    ‘zero’, ini_method=‘intensity’, 
    adaptive_rms_box=True, adaptive_thresh=150, 
    rms_box_bright=(60,15), group_by_isl=False, 
    group_tol=10.0, output_opts=True, 
    output_all=False, atrous_do=True, 
    atrous_jmax=4, flagging_opts=True, 
    flag_maxsize_fwhm=0.5, advanced_opts=True, 
    blank_limit=None, frequency=restfreq)
\end{verbatim}
\hspace{\parindent}where \texttt{image} is the primary beam corrected image, \texttt{app\_image} is 

the image without a primary beam  correction applied and \texttt{restfreq} 

is the frequency of the survey in Hz.

\begin{itemize}
    \item Next, we inject the brightest ($\geq150\sigma_{thresh}$) sources back into the \texttt{resid\_map\_1} image to create \texttt{resid\_map\_with\_bright}. This is done because \textsc{PyBDSF} uses a smaller sliding box size around bright sources in order to better characterise the elevated noise in these regions. By injecting the sources back into  \texttt{resid\_map\_1} we ensure that subsequent \textsc{PyBDSF} runs use the same sliding box behaviour as previous runs.
    \item Then, we run \textsc{PyBDSF} as above on \texttt{resid\_map\_with\_bright} and use this to calculate an RMS map (\texttt{rms\_map\_deep}) which is deeper than that obtained from the initial \textsc{PyBDSF} run because the source density of \texttt{resid\_map\_with\_bright} is far lower than our original image.
    \item To produce a deeper and more complete catalogue we run \textsc{PyBDSF} over the original image using the same parameters as above, except passing it the RMS map (\texttt{rms\_map\_deep}) directly. This creates a residual image which we denote as \texttt{resid\_map\_deep}. The complete configuration run is:
\end{itemize}
    
\begin{verbatim}
bdsf.process image(image, detection_image=
    app_image, thresh_isl=3.0, thresh_pix=4.0,
    rms_box=(150,15), rms_map=True, mean_map=
    ‘zero’, ini_method=‘intensity’,
    adaptive_rms_box=True, adaptive_thresh=150, 
    rms_box_bright=(60,15), group_by_isl=False, 
    group_tol=10.0, output_opts=True,
    output_all=False, atrous_do=True, 
    atrous_jmax=4, flagging_opts=True, 
    flag_maxsize_fwhm=0.5, advanced_opts=True, 
    blank_limit=None, frequency=restfreq, 
    rmsmean_map_filename=[meanmap, rms_map_deep], 
    rmsmean_map_filename_det=[meanmap, 
    rms_map_deep_app])
\end{verbatim}

\begin{itemize}
    \item Finally, as a visual inspection of the \textsc{PyBDSF} residual maps indicates that complex structures are not always adequately modeled, we run \textsc{PyBDSF} on \texttt{resid\_map\_deep} to pick up emission which may have been missed by \textsc{PyBDSF} in the previous run
    using the configuration:
\end{itemize}

\begin{verbatim}
bdsf.process image(resid_map_deep, detection_image=
    resid_map_deep_app, thresh_isl=3.0, thresh_pix=
    10.0, rms_box=(150,15), rms_map=True, 
    mean_map=‘zero’, ini_method=‘intensity’, 
    adaptive_rms_box=True, adaptive_thresh=150, 
    rms_box_bright=(60,15), group_by_isl=False, 
    group_tol=10.0, output_opts=True, 
    output_all=False, flagging_opts=True, 
    flag_maxsize_fwhm=0.5, advanced_opts=True, 
    blank_limit=None, frequency=restfrq, 
    rmsmean_map_filename=[meanmap, rms_map_deep], 
    rmsmean_map_filename_det=[meanmap, 
    rms_map_deep_app], flag_maxsize_bm=100)
\end{verbatim}

\begin{itemize}
    \item We add these sources to the final source and Gaussian component catalogues. We also update the \texttt{resid\_map\_deep} map accordingly so that it reflects the final residual image.
\end{itemize}

 \end{appendix}

\end{document}